\documentclass[aos,preprint]{imsart}

\RequirePackage[OT1]{fontenc}
\RequirePackage{amsthm,amsmath}

\usepackage{enumitem}
\usepackage{amsmath,amsfonts,amsthm,amssymb,dsfont,bbm}
\usepackage{geometry} 
\usepackage{xr-hyper}
\RequirePackage[colorlinks,citecolor=blue,urlcolor=blue]{hyperref}
\allowdisplaybreaks[4]

\usepackage{standalone}
\arxiv{arXiv:0000.0000}

\startlocaldefs
\numberwithin{equation}{section}
\theoremstyle{plain}
\newtheorem{theo}{Theorem}[section]
\newtheorem{rem}{Remark}[section]

\newtheorem{Example}{Example}[section]

 \newcommand{\ind}{\mathbbm{1}}
 \def\proof{\noindent {\bf Proof.}\ }
\def\endproof{{\mbox{}\nolinebreak\hfill\rule{2mm}{2mm}\par\medbreak} }
\def\E{{\mathbb{E}}}
\def\P{{\mathbb{P}}}
\def\F{{\mathcal F}}
\def \Tr{w}

\newlist{assum}{enumerate}{2}
\setlist[assum,1]{label=(A\arabic*),labelindent=*,resume}
\setlist[assum,2]{label=(A\arabic{assumi}\alph*),leftmargin=0pt,labelindent=0pt}
\endlocaldefs

\begin{document}

\begin{frontmatter}
\title{Nonparametric covariate-adjusted response-adaptive design based on a functional urn model}
\runtitle{CARA Functional Urn Design}

\begin{aug}
\author{\fnms{Giacomo} \snm{Aletti}\thanksref{m1}\ead[label=e1]{giacomo.aletti@unimi.it}},
\author{\fnms{Andrea} \snm{Ghiglietti}\thanksref{m2}\ead[label=e2]{andrea.ghiglietti@unicatt.it}}\\
\and
\author{\fnms{William} \snm{F. Rosenberger}\thanksref{m3}\ead[label=e3]{wrosenbe@gmu.edu}}

\runauthor{Aletti \textit{et. al.}}

\affiliation{ADAMSS Center and Universit\`{a} degli Studi di Milano\thanksmark{m1},\\
Universit\`{a} Cattolica del Sacro Cuore\thanksmark{m2},\\
George Mason University\thanksmark{m3}}

\address{Department of Environmental Science and Policy\\
Universit\`{a} degli Studi di Milano\\
via Saldini 50\\
20133 Milan, Italy\\
\printead{e1}}

\address{Department of Statistical Science\\
Universit\`{a} Cattolica del Sacro Cuore\\
via Largo Gemelli 1\\
20123 Milan, Italy\\
\printead{e2}}

\address{Department of Statistics\\
George Mason University\\
4400 University Drive MS4A7\\
Fairfax, VA 22030 USA\\
\printead{e3}}

\end{aug}

\begin{abstract}
In this paper we propose a general class of covariate-adjusted response-adaptive (CARA) designs
based on a new functional urn model.
We prove strong consistency concerning the functional urn proportion and
the proportion of subjects assigned to the treatment groups, in the whole study and for each covariate profile,
allowing the distribution of the responses conditioned on covariates to be estimated nonparametrically.
In addition, we establish joint central limit theorems for the above quantities and the sufficient statistics of features of interest,
which allow to construct procedures to make inference on the conditional response distributions.
These results are then applied to typical situations concerning Gaussian and binary responses.
\end{abstract}

\begin{keyword}[class=MSC]
\kwd[Primary ]{62L20}
\kwd{60F05}\kwd{62E20}\kwd{62L05}
\kwd[; secondary ]{62F12}
\kwd{62P10}
\end{keyword}

\begin{keyword}
\kwd{Clinical trials}
\kwd{covariate-adjusted analysis}
\kwd{inference}
\kwd{large sample theory}
\kwd{personalized medicine}
\kwd{randomization}
\end{keyword}

\end{frontmatter}

\section{Introduction}\label{introduction}

\subsection{Adaptive designs in clinical trials}

The scientific validation of new treatments or therapies in medicine is
typically the result of careful controlled randomized clinical trials.
The standard methodology is to sequentially assign the patients to the treatment groups and
collect the corresponding responses for statistical inference.
The design of such experiments typically involves several aspects,
such as ethical objectives, reduction of costs, and inferential properties.
Initially, randomized adaptive procedures were used to increase the balance among the treatment groups
in order to achieve an unbiased comparison (e.g. see \cite{Ros,RosLac}).
However, balance does not ensure efficiency or good ethical properties, except in very particular circumstances (see~\cite{RosSve}).
Hence, new procedures, called response-adaptive, have been considered that use the accrued information on previous subjects' response
to treatments to skew the probabilities of assignment away from $1/2$ towards specific target values.
An exhaustive review on response-adaptive procedures can be found for instance in~\cite{AtkBis14,HuRos}.
These designs are typically constructed to satisfy
certain optimality criteria related to their performance,
with respect to ethical aspects, such as minimizing the expected number of failures, or statistical properties,
such as maximizing the power of the test.
Specifically, a standard methodology is to consider a desired asymptotic allocation proportion $\mathbf{\rho}$
depending on the response distributions that satisfies those optimality criteria.
In a common situation, the response distributions depend on a vector of parameters $\mathbf{\theta}\in\Theta$,
and the target allocation is defined as a function $\rho:\Theta\rightarrow \mathcal{S}$,
where $\mathcal{S}$ is the simplex of dimension equal to the number of treatments.
Thus, the response-adaptive design is constructed such that the proportion of subjects assigned to the treatments
asymptotically target the desired target proportion, i.e. $\mathbf{N}_n/n\stackrel{a.s.}{\rightarrow}\rho(\mathbf{\theta})$.

When the information on significant covariates is available,
adaptive designs based solely on the patient's response to treatments are inadequate to implement the randomized assignments.
For instance, when a covariate has a strong influence on the response to a treatment,
it may be inappropriate to use responses observed by subjects with a specific covariate profile to determine
patients' allocation probability with a different profile.
As described in~\cite{RosSve}, there are multiple ways for taking into account the effect of covariates in clinical trials and,
in general, there is no agreement about how to implement such designs and what should be the main purpose of these procedures.

\subsection{Covariate-adjusted randomization designs}
A natural way to incorporate covariates in randomized procedures is to adopt stratification to force balance on certain important covariates.
In fact, although randomization reduces the probability that the presence of the covariates are strongly different in the treatment groups
(see~\cite{RosLac}), perfect balance is reached only asymptotically and for small samples a significant imbalance can easily occur.
Hence, a standard methodology is to stratify on few important known covariates, then to use restricted
randomization within each stratum, and finally to let randomization handle the less influential and unknown covariates.
Since this approach is possible only for a small number of covariates,
several procedures have been proposed to individualize which covariates should be considered in the trial and how to implement the sequential assignments.
An early work on these covariate-adjusted randomization designs is represented by the
deterministic procedure proposed in~\cite{Tav} to minimize imbalances on strata
and its extensions that include randomization:
the biased coin design of~\cite{PocSim}
and the marginal urn design formulated by~\cite{Wei78}.
In this context, different types of imbalances have been controlled in the covariate-adaptive designs proposed in~\cite{HuHu},
by using the positive recurrence of the Markov process of the within-stratum imbalances.
Moreover, a wide class of covariate-adaptive designs aimed at balancing the allocations has been recently presented in~\cite{BaldiZag17},
in which the main asymptotic properties have been established.
However, as shown in~\cite{RosSve}, balance does not guarantee either to have a design with good statistical properties or
to assign more patients to the superior treatment.
An alternative approach was proposed in~\cite{Atk82}
to incorporate treatment-by-covariate interactions and continuous covariates.
Specifically, the subjects are assigned to treatment groups in order to minimize
the variance of the treatment responses conditionally on their covariate profiles.
This procedure is based on the $D_A$-optimality criterion with linear models and, unlike the previous designs,
it performs well also for correlated covariates.
However, the minimum variance of the treatment responses can be obtained through balancing covariates in marginal strata only in linear models with
homoscedastic errors, but it is not valid in nonlinear models
(see e.g.
logistic regression~\cite{KalHar}) or when errors present a general covariance structure.
Hence, information on the covariates may not be sufficient to achieve efficiency or ethical goals, and
using the responses to treatments in the allocation phase can be essential to this task.
For this reason, adaptive procedures that use ether the covariate profile of the subjects and
the performances of the treatments have been considered in literature under the name
covariate-adjusted response-adaptive (CARA) designs (see~\cite{RosVidAga}).

\subsection{CARA designs}
In general, a CARA design is defined as a procedure that randomly assigns subjects to the treatment groups with a probability that depends on
their own covariate profiles and on the previous patients' covariates, allocations and responses.
The literature on this class of designs is not very long and
early steps in this context can be found in~\cite{Kad} and~\cite{RosVidAga}.

Concerning binary responses and polytomous covariates, a relevant CARA design based on the randomized play-the-winner rule has been proposed in~\cite{BanBis99}.
The case of binary responses and two competing treatments has been also considered in~\cite{RosVidAga} for different types of covariates.
The allocation rule proposed in~\cite{RosVidAga} uses a suitable mapping based on a logistic regression model
that implements the interaction among covariates and responses to treatments.
Specifically, each subject is assigned according to the odds ratio comparing treatments in correspondence of his own covariate profile.
The main properties of this adaptive allocation rule has been investigated through simulation,
highlighting a significant reduction of the expected treatment failures.
Nevertheless, theoretical results on the design performances have not been derived in~\cite{RosVidAga}.
A two-stage CARA design for binary responses based on logistic regression model was implemented in~\cite{BanBisBha},
in which the patients are assigned initially using a restricted procedure to compute the adaptive parameter estimators and
then using a probability that depends on such estimators and the corresponding covariate profile.

The case of continuous responses and two treatments has been considered in~\cite{BanBis01},
in which an adaptive design with limiting allocation proportion has been proposed
by using a linear model to incorporate covariate information.
However, the probability of assigning the next patient does not depend on its covariate profile,
and hence the design proposed in~\cite{BanBis01} cannot be included in the classical CARA framework.
The procedure in~\cite{BanBis01} has been improved in~\cite{AtkBis05a}
by considering an adaptive biased-coin design for normal responses based on a generalized $D_A$-optimal criterion
that takes into account both statistical and ethical purposes.
Although the performances of these allocation rules have been studied through simulation,
their theoretical properties have not been proved.

Ideally, the analysis of the ethical and inferential properties of the experimental designs should be based on theoretical results concerning
the asymptotic behavior of the allocation proportion and adaptive estimators,
and none of the previous work on CARA designs is able to provide such results.
In fact, since the allocation and the estimation process depend on both the responses and the covariates,
CARA designs are very complex to be formulated in a rigorous mathematical setting.
Two papers, in particular, formalize CARA in a rigorous mathematical framework. The first of these is the groundbreaking paper
of \cite{ZhaHuCheCha07},
in which consistency and second-order asymptotic results concerning both adaptive estimators and allocation proportions have been proved
for a very wide class of CARA designs.
In the second \cite{BaldiZag}, compound optimal design theory was used to find target allocations of interest, and these target allocations are attained using an accelerated biased coin design.

The procedures considered in~\cite{ZhaHuCheCha07} can be interpreted as generalized biased coin designs in which the probability of allocation
is given by a known target function evaluated at the adaptive estimators
of a finite number of parameters related to the responses means conditioned on the covariates.
For this reason, this design has been applied to generalized linear models for different types of responses and covariates (discrete and continuous)
and for more than two treatments.
Some recent papers have extended the class of CARA designs presented in \cite{ZhaHuCheCha07} in order to improve its inferential properties.
For instance, the class of designs proposed in~\cite{Zhu} allows inference also for common parameters in the response distributions.
The unified family of designs presented in~\cite{HuZhuHu} takes into account both efficiency and medical ethics.
The distribution of the parameters estimators in reduced generalized linear models established in~\cite{CheungZhangHuChan}
allows inference for separately testing the main effects, the covariate effects or their intersections.
One of the aims of this paper is to construct a new framework for CARA designs in which the probability of allocation
may depend by nonparametric or semi-parametric estimates of the generic conditional response distribution.
To this purpose, the proposed design is based on the other very popular class of randomized procedures: urn models.

\subsection{Urn models}
The history of urn models as probabilistic tools to describe random phenomena is very long and deep
in several fields of scientific research (e.g.~\cite{Beg} in economics,~\cite{BenSchTar} in genetics,~\cite{AleCriGhi16,AleGhi16} in network analysis).
Starting from the P\'{o}lya urn proposed in~\cite{EggPol} to model contagious disease,
many variants have been considered
(see e.g.~\cite{ale-ghi-pag,AthKar68,BaiHu99,ghi-vid-ros,MayFlo09,ZhaHuChe,ZhaHuCheCha11}).
In this paper, we focus on the broad class of urn schemes known as generalized P\'{o}lya urn (GPU);
its asymptotic behavior has been the objects of several important works:
starting from the asymptotic results proved in~\cite{AthKar68} by embedding the urn process in a continuous-time branching process,
other significant theoretical results have been derived, for instance in~\cite{BaiHu99,BaiHu05,BaiHuZha,Jan,Smy,ZhaHuChe,ZhaHuCheCha11}.
Concerning applications, urn models have known a great popularity as adaptive designs in clinical trials for several reasons:
(i) the process of colors sampled from the urn represents a natural way to model the sequential randomized assignments of subjects to treatments groups,
(ii) the quantity of balls replaced in the urn may depend on previous patients' information
so that different types of adaptive procedures can be constructed,
(iii) their updating rule composed by extractions and replacements is intuitive and easy to be implemented by clinicians.
In clinical trials, urn models are mostly adopted as response-adaptive designs and covariates are not considered.
For a review of urn models adopted as response-adaptive procedures see~\cite{DurFloLi98,MayFlo09,HuRos}.

The response-adaptive GPU design presented in~\cite{ZhaHuChe} is the following:
consider an urn containing balls of $d$ colors, each one associated with a specific treatment;
the composition of the urn at time $n\geq0$ is represented by the vector $\mathbb{Y}_n=(\mathbb{Y}_n^1,..,\mathbb{Y}_n^d)^\top\in\mathbb{R}_+^d$.
At any time $n\geq1$, a subject enters the trial, a ball is sampled at random from the urn and its color is observed;
formally, let $\mathbb{X}_n\in\{0,1\}^d\cap\mathcal{S}$ represent the color sampled at time $n$:
for each $j\in\{1,..,d\}$, $\mathbb{X}^j_n=1$ indicates that the sampled ball is of color $j$, $\mathbb{X}^j_n=0$ otherwise.
Then, letting $k\in\{1,..,d\}$ be the sampled color,
the subject is assigned to treatment $k$ and a response ${\xi}^k_n$ is observed.
The potential responses $\{\xi^k_n;n\geq1\}$, $k\in\{1,..,d\}$,
are defined as $d$ independent sequences of identically distributed (i.i.d.) random variables.
Moreover, for any $i,j\in\{1,..,d\}$, the model uses a function $u^{ij}:S^j\rightarrow \mathbb{R}_+$ ($S^j$ indicates the support of ${\xi}^j_n$)
to convert the responses into the reinforcements;
then, the sampled ball of color $k$ is returned to the urn together with $u^{ik}({\xi}^{k}_n)$ new balls of color $i\in\{1,..,d\}$.
Formally, denoting $\mathbb{D}_n$ the $d\times d$ replacement matrix defined as $\mathbb{D}^{ij}_n:=u^{ij}({\xi}^j_n)$ for any $i,j\in\{1,..,d\}$,
the urn is updated as follows
\begin{equation}\label{eq:urn_scheme_n_T}
\mathbb{Y}_n\ =\ \mathbb{Y}_{n-1}\ +\ \mathbb{D}_n\mathbb{X}_n.
\end{equation}
Note that $\mathbb{D}_n^{ik}$ can be computed only if the subject receives treatment $k$,
because ${\xi}^k_n$ is needed to obtain $\mathbb{D}_n^{ik}=u^{ik}({\xi}^k_n)$.
However, the updating rule expressed in~\eqref{eq:urn_scheme_n_T}
does not require the value of $\mathbb{D}_n^{ik}$ if the subject receives a treatment different than $k$,
since in that case $\mathbb{X}^k_n=0$.

In the GPU model, it is well known that the asymptotic behavior of the urn proportion is related
to the limit of the conditional expectation of the replacement matrices, i.e. $\mathbb{H}:=\lim_{n\rightarrow\infty}\E[\mathbb{D}_n|\mathcal{F}_{n-1}]$
a.s. (e.g. \cite{AthKar68,BaiHu99,BaiHu05,BaiHuZha,ZhaHuChe,ZhaHuCheCha11}),
where $\mathcal{F}_{n-1}$ indicates the quantities observed up to time $(n-1)$.
Specifically, under suitable conditions, we have that $\mathbb{Y}_n/\Tr(\mathbb{Y}_n)\stackrel{a.s.}{\rightarrow}\mathbb{V}$,
where $\Tr(\mathbb{Y}_n)=\sum_{j=1}^d\mathbb{Y}^j_n$ and
$\mathbb{V}\in\mathcal{S}$ is the right eigenvector of $\mathbb{H}$ associated with its maximum eigenvalue and such that $\Tr(\mathbb{V})=1$.
Hence, in order to target a specific proportion $\rho(\mathbf{\theta})$, one needs to define appropriately the replacement matrices $\mathbb{D}_n$
such that $\mathbb{H}$ guarantees $\rho(\mathbf{\theta})\equiv\mathbb{V}$.
To this end, the functions $u^{ij}$ should depend on the parameters $\mathbf{\theta}$, which are usually unknown in practice.
Hence, in~\cite{ZhaHuChe} the parameter $\mathbf{\theta}$ is replaced in $u^{ij}$ by the adaptive estimator
$\mathbf{\hat{\theta}}_{n-1}$ computed with the information available up to time $(n-1)$ obtaining $\hat{u}^{ij}$,
so that the replacement matrix is represented by
$\mathbb{D}^{ij}_n=\hat{u}^{ij}(\xi^j_n)$ for $i,j\in\{1,..,d\}$.
This model proposed in~\cite{ZhaHuChe} is called a \emph{sequential estimation-adjusted urn} model (SEU) and
it has been proved that the model targets any desired limiting proportion $\rho(\theta)$.

\subsection{Aim and organization of the paper}
Although the SEU model represents a very powerful urn design,
as with most response-adaptive procedures, it does not incorporate the covariates in the randomization process.
Hence, the probability of allocating a subject to a treatment group is independent of his or her covariate profile;
moreover, all patients are asymptotically assigned with the same target probability $\rho(\theta)$, regardless of their covariate profiles.
In this paper, we want to extend the SEU model by introducing information on the covariates in the urn scheme, so obtaining a CARA urn design.
Analogously to the CARA coin designs described in~\cite{ZhaHuCheCha07},
where the covariates are considered in the trial,
each subject is assigned with a probability that depends on his or her own covariate profile.
Formally, let $\tau$ be the covariate space, which could be finite, countable, or continuous. Conditionally on the covariate profile $t\in\tau$,
we consider different response distributions $\pi^1_t,..,\pi_t^d$ and the corresponding parameters $\mathbf{\theta}_t$.
Thus, in this framework the main goal of the design is to
asymptotically assign all the subjects with covariate profile $t$ with a desired probability $\rho(\mathbf{\theta}_t)$,
i.e. the design targets the desired functional allocation $\rho(t)$.
We will see that the model presented in this paper achieves this goal.
It is worth noticing that in the design proposed in~\cite{ZhaHuCheCha07},
$\mathbf{\theta}_t$ characterizes the means of the responses conditionally on $t$
and it is defined as $\mathbf{\theta}_t=f(\mathbf{\beta},t)$, where $f$ is a known function and $\mathbf{\beta}$ is a finite set of parameters.
As a consequence, in~\cite{ZhaHuCheCha07} the target function $\rho(t)$ actually depends on a finite number of unknowns represented by $\beta$.
The framework considered in this paper is different since
$\mathbf{\theta}_t$ may represent general features of the conditional response distributions
and it can be estimated nonparametrically; the target function $\rho(t)$ depends on an infinite number of unknowns.
Another difference is that in the existing literature
the probability distribution of the covariates is the same for all the patients,
while in this paper this distribution is allowed to be adaptively modified by the experimenter
using the information collected during the trial.

The CARA design we propose consists of a functional urn model
in which the urn composition is a $d$-dimensional multivariate function of the covariates.
Each subject is assigned to the treatment group according to the color sampled from the urn identified by his or her covariate profile.
After any allocation, the entire functional urn composition is updated,
even if only the response associated with the patient's covariate profile has been observed.
For this reason, a crucial point is the definition of the functional objects $\mathbf{X}_n$ and $D_n$
that extend the multivariate objects $\mathbb{X}_n$ and $\mathbb{D}_n$ in the updating rule~\eqref{eq:urn_scheme_n_T}.

In this paper we establish first and second-order asymptotic results concerning the following quantities:
\begin{itemize}
\item[(i)] the probability of allocation of the subjects for each covariate profile, $\mathbf{Z}_n(t)$;
\item[(ii)] the proportion of subjects assigned to the treatment groups for each covariate profile, $\mathbf{N}_{t,n}/\Tr(\mathbf{N}_{t,n})$;
\item[(iii)] the proportion of subjects assigned to the treatment groups in the trial, $\mathbf{N}_n/n$;
\item[(iv)] the adaptive estimators $\hat{\theta}_{t,n}$ of features of interest $\theta_t$ related with the
distribution of the treatments responses conditionally on each covariate profiles
(required for the inference based on covariate-stratification approach);
\item[(v)] the adaptive estimators $\hat{\beta}_{n}$ of features of interest $\beta$ related with
the entire family of response distributions conditionally on the covariates
(required for the inference based on covariate-adjusted approach);
\end{itemize}
In particular, we prove strong consistency of the above quantities,
allowing the distribution of the responses conditioned on the covariates to be estimated nonparametrically.
In addition, we establish joint central limit theorems (CLTs)
which provide the essential probabilistic tools to construct inferential procedures on conditional response distributions.
These results are then applied to typical situations concerning Gaussian and binary responses.

In Section~\ref{section_model} we present the CARA urn design based on the functional urn model.
Section~\ref{section_assumptions_results} is concerned with assumptions and main results.
In Section~\ref{section_examples_linear_regression} and~\ref{section_examples_logistic_regression}
the CARA urn design is applied to different practical scenarios
and several response distributions are considered.
Concluding remarks and future developments are discussed in Section~\ref{section_conclusions}.
Technical details, including proofs of the theorems, are presented in a supplement \cite{AleSupp}.

\section{The functional urn model for CARA designs}\label{section_model}

In this section we describe the design based on a functional urn model.
We start by defining the quantities related to the process of subjects that sequentially enter the trial.

\subsection{Notation}\label{subsection_notaion}
Consider a trial in which patients are sequentially and randomly assigned to $d\geq2$ treatments.
For any $n\geq1$, let $\mathbf{\bar{X}}_n\in\{0,1\}^d\cap\mathcal{S}$ represent the treatment assigned to subject $n$:
for any $j\in\{1,..,d\}$, $\bar{X}^j_n=1$ indicates that the assigned treatment is $j$, $\bar{X}^j_n=0$ otherwise.
Each subject $n$ is identified by a vector $(T_n,\xi^1_n,..,\xi^d_n)\in(\tau\times S^1\times..\times S^d)$,
in which $T_n$ indicates his or her covariate profile and,
for any $j\in\{1,..,d\}$, $\xi^j_n$ indicates the patient's potential response to treatment $j$.
In general, the covariate profiles of the patients $\{T_n;n\geq1\}$ are
a sequence of independent but nonidentically distributed random variables, whose
distributions can adaptively depend on the information collected during the trial:
the covariate profiles, the allocations and the responses. This allows a generalization of the typical CARA design, by
incorporating a probabilistic mechanism to select patients with a particular covariate profile if patient selection is also
a random process.
This includes as a special case the typical assumption of $\{T_n;n\geq1\}$ as i.i.d. random variables,
in which the information collected in the trial does not affect the choice of the future covariate profiles. This represents
the standard clinical trial where the clinician has no control over the patient recruitment process.
We will denote by $\mu_{n-1}$ the probability distribution of $T_{n}$ conditioned on
$\mathcal{F}_{n-1}$, i.e. the information collected up to time $(n-1)$.

Since any subject receives one treatment, for any $n\geq1$ only one value among $\{\xi^1_n,..,\xi^d_n\}$ can be observed during the trial
and we will denote it $\bar{\xi}_n$.
For each $j\in\{1,..,d\}$, we assume that $\{\xi^j_n;n\geq1\}$ is a sequence of independent random variables
whose distribution depends on the sequence of covariate profiles $\{T_n;n\geq1\}$.
Specifically, we define a family of probability distributions $\{\pi_{t}^j;t\in\tau\}$,
where each one represents the distribution of $\xi^j_n$ conditioned on the event $\{T_n=t\}$:
$\pi_{t}^j$ indicates the probability law of the response to treatment $j$ observed from
a subject whose covariate profile is equal to $t$. Note that $\{T_n=t\}$ could have measure zero.
In the paper, we will also use the corresponding families of cumulative distribution functions (CDFs) $\{F_{t}^j;t\in\tau\}$
and the families of quantile functions (QFs) $\{Q_{t}^j;t\in\tau\}$.

\subsection{The model}\label{subsection_model}
For any $n\geq0$, let $\mathbf{Y}_n=(Y_n^1,..,Y_n^d)^\top$ be a $d$-dimensional vector of nonnegative bounded functions
and let $\mathbf{Z}_n=\mathbf{Y}_n/\Tr(\mathbf{Y}_n)$, where $\Tr(\mathbf{Y}_n):=\sum_{j=1}^dY_n^j$.
For any $t\in\tau$, $\mathbf{Y}_n(t)\in(0,1)^d$ represents an urn containing $Y^j_n(t)$ balls of color $j\in\{1,..,d\}$
and $\mathbf{Z}_n(t)\in\mathcal{S}$ indicates the proportions of the colors in the urn at time $n$.
To avoid inessential complications, we consider a uniform initial composition $\mathbf{Y}_0=\mathbf{1}$.
For any $n\geq1$, let $\mathcal{F}_{n-1}$ be the $\sigma$-algebra composed by the information related with the first $(n-1)$ patients,
i.e. their covariate profiles, allocations and responses:
\begin{equation}\label{def:sigma_algebra_n-1}
\mathcal{F}_{n-1}\ :=\ \sigma\left(T_1,\mathbf{\bar{X}}_1,\bar{\xi}_1,..,T_{n-1},\mathbf{\bar{X}}_{n-1},\bar{\xi}_{n-1}\right).
\end{equation}
When subject $n$ enters the trial, his or her covariate profile $T_n$ is observed, and we now operate on the conditioning set
$\{\mathcal{F}_{n-1},T_n\}$, consistent with the definition of CARA designs found in \cite{HuRos}.
Then, a ball is sampled at random from the urn identified by $T_n$;
i.e., with proportions $$\mathbf{Z}_{n-1}(T_n) =\left(Z_{n-1}^1(T_n), .., Z_{n-1}^d(T_n)\right)^\top$$ and its color is observed;
thus, the subject $n$ receives the treatment associated with the sampled color and a response $\bar{\xi}_n$ is collected.
In order to update the functional urn, we construct a weighting function $\mathbf{X}_n$ and a functional replacement matrix $D_n$
that extend $\mathbb{X}_n$ and $\mathbb{D}_n$ in the classical updating rule; see~\eqref{eq:urn_scheme_n_T}.

First, we define the weighting function $\mathbf{X}_n$.
Let $U_n$ be a uniform (0,1) random variable independent of $\mathcal{F}_{n-1}$ and $T_n$, and define, for any $t\in\tau$ and
for any $j\in\{1,..,d\}$
\begin{equation}\label{def:X_breve_n_statistical_part}
\breve{X}_n^j(t):=\ind_{\{\sum_{i=1}^{j-1}Z_{n-1}^{i}(t)<U_n\leq\sum_{i=1}^{j}Z_{n-1}^i(t)\}},
\end{equation}
where we use the convention $\sum_{i=1}^{0}(\cdot)=0$.
Notice that $\mathbf{\breve{X}}_n(t)$ represents the color of the ball that would be sampled
if we used the urn identified by $t$; i.e. if the covariate profile $T_n$ were equal to $t$.
Thus, $\mathbf{\bar{X}}_n:=\mathbf{\breve{X}}_n(T_n)$ represents the color actually sampled from the functional urn at time $n$:
for each $j\in\{1,..,d\}$, $\bar{X}^j_n=1$ indicates that the sampled ball is of color $j$, $\bar{X}^j_n=0$ otherwise.
Hence, conditionally on $\mathcal{F}_{n-1}$ and $T_n$, $\bar{X}^j_n$ is Bernoulli distributed with parameter $Z^j_{n-1}(T_n)$.
Since $\mathbf{\breve{X}}_n(t)$ models the color hypothetically sampled from the urn identified by $t$ (i.e. $\mathbf{Y}_{n-1}(t)$),
we define the weighting function $\mathbf{X}_n$ as the expected value of $\mathbf{\breve{X}}_n$
conditioned on the information of the color sampled from the urn identified by $T_n$, i.e.
\begin{equation}\label{def:X_n_statistical_part}
\mathbf{X}_n\ :=\ \E\left[\ \mathbf{\breve{X}}_n\ |\ \mathcal{F}_{n-1},T_n,\mathbf{\bar{X}}_n\ \right].
\end{equation}
An analytic expression of $\mathbf{X}_n$ derived from~\eqref{def:X_n_statistical_part} is provided in~\eqref{def:X_n_statistical_part_simply} in the supplement \cite{AleSupp}.
Note that by~\eqref{def:X_breve_n_statistical_part} and~\eqref{def:X_n_statistical_part} we have $\mathbf{X}_n(t)\in \mathcal{S}$ for any $t\in\tau$,
since $\mathbf{X}_n(t)\in[0,1]^d$ and
$$\Tr(\mathbf{X}_n)=\Tr\left(\E\left[\mathbf{\breve{X}}_n|\mathcal{F}_{n-1},T_n,\mathbf{\bar{X}}_n\right]\right)=
\E\left[\Tr(\mathbf{\breve{X}}_n)|\mathcal{F}_{n-1},T_n,\mathbf{\bar{X}}_n\right]=1.$$
Moreover, we also have that
\begin{equation}\label{eq:expetcted_X_equal_Z}
\E\left[\mathbf{X}_n|\mathcal{F}_{n-1}\right]\ =\ \mathbf{Z}_n,
\end{equation}
since by the law of total expectation $\E\left[\mathbf{X}_n|\mathcal{F}_{n-1}\right]=\E[\mathbf{\breve{X}}_n|\mathcal{F}_{n-1}]=\mathbf{Z}_{n-1}$.

We now define the functional replacement matrix $D_n$.
First, for any $i,j\in\{1,..,d\}$ and $t\in\tau$, let $u^{ij}_t:S^j\rightarrow \mathbb{R}_+$ be a function that
converts the responses into the reinforcements.
Analogously to the classical updating rule~\eqref{eq:urn_scheme_n_T},
for any $t\in\tau$, the urn identified by $t$ should be ideally updated
by $u^{ij}_t(\xi_n^j)$ balls of color $i\in\{1,..,d\}$ when treatment $j$ is assigned,
where $\xi_n^j$ represents the response observed from a subject with covariate profile $t$.
Hence, the urn identified by $t$ should be updated
by $u^{ij}_t(W_t^j)$ balls of color $i\in\{1,..,d\}$,
where $W_t^j$ represents a random variable with probability distribution $\pi^j_{t}$ and that,
conditionally on $\mathcal{F}_{n-1}$ and $T_n$, is independent of $\mathbf{\bar{X}}_n$.
This can be equivalently formalized by introducing a uniform (0,1) random variable $V_n$ independent of $U_n$, $T_n$ and $\mathcal{F}_{n-1}$,
and defining $W_t^j=Q^j_{t}(V_n)$ for any $j\in\{1,..,d\}$,
where we recall that $Q^j_{t}$ is the QF associated with the probability distribution $\pi^j_{t}$.
In fact, by definition, we have that $Q^j_{t}(V_n)\sim \pi_{t}^j$ when $V_n\sim U(0,1)$.
Thus, the replacements in the urn identified by $t\in\tau$ should be defined by the following random matrix:
\begin{equation}\label{def:D_breve_n_statistical_part}
\breve{D}_n^{ij}(t)\ :=\ u^{ij}_t(\ Q^j_{t}(V_n)\ ),\ \ \ \forall\ i,j\in\{1,..,d\}.
\end{equation}
However, when the subject $n$ with covariate profile $T_n$ is assigned to a treatment $j$ and the response $\xi_n^{j}$ is observed,
we can only compute $\bar{D}^{ij}_n:=u^{ij}_{T_n}(\xi_n^{j})$, that corresponds to $\breve{D}_n^{ij}(T_n)$.
Nevertheless, the response $\xi_n^{j}$, associated with the covariate profile $T_n$,
contains the information on the quantile $V_n$ that can be taken into account to update all the urns $t\in\tau$ using~\eqref{def:D_breve_n_statistical_part}.
Specifically, the $d\times d$ replacement matrix of bounded functions
should be defined as the expected value of the potential replacement matrix $\breve{D}_n$ for the urn $t$,
conditionally on the information of the response observed to treatment $k\in\{1,..,d\}$, that we call $\bar{\xi}_n=\xi^{k}_{n}$,
from a subject with covariate profile $T_n$, i.e.
\begin{equation}\label{def:D_n_statistical_part}
D^{*}_n\ :=\ \E\left[\ \breve{D}_n\ |\ T_n,\mathbf{\bar{X}}_n,\bar{\xi}_n\ \right].
\end{equation}
An explicit expression of~\eqref{def:D_n_statistical_part} can be derived as follows:
for any $s\in\tau$, $k\in\{1,..,d\}$ and $y\in S^k$,
\begin{equation}\label{def:D_n_statistical_part_simply}
\begin{aligned}
D_n^{*ij}&=\ \E\left[\ \breve{D}^{ij}_n\ |\ \{T_n=s\},\{\bar{X}_n^k=1\},\{\bar{\xi}_n=y\}\ \right]\\
&=\ \E\left[\ u^{ij}_t(\ Q^j_{t}(V_n)\ )\ |\ \{V_n\in (Q^k_{s})^{-1}(y)\}\ \right],
\end{aligned}
\end{equation}
where
\begin{equation*}
(Q^k_{s})^{-1}(y)\ :=\ \left\{\ v\in(0,1)\ :\ Q^k_{s}(v)=y\ \right\},
\end{equation*}
and we recall that $Q^k_{s}$ is the QF associated with the probability distribution $\pi^k_{s}$.
Note from~\eqref{def:D_n_statistical_part} that $D^{*}_n$ depends on quantities that are unknown at time $n$.
Specifically, the expression in~\eqref{def:D_n_statistical_part} contains the conditional QFs $Q^j_{t}$ and $Q^k_{s}$;
moreover, as mentioned in Section~\ref{introduction},
$u^{ij}_t$ typically depends on the response distribution $\pi^j_{t}$
in order to obtain some desired asymptotic properties from the design.
Hence, since the conditional distributions $\pi^j_{t}$ are typically unknown,
we compute the corresponding functional estimators $\hat{u}^{ij}_t$, $\hat{Q}^j_{t}$ and $\hat{Q}^k_{s}$
by using the information related with the first $(n-1)$ subjects.
Thus, the $d\times d$ replacement matrix $D_n$ of bounded functions is defined,
on the sets $\{T_n=s\},\{\bar{X}_n^k=1\},\{\xi^{k}_n=y\}$, as follows:
\begin{equation}\label{def:D_n_statistical_part_true}
D_n^{ij}\ :=\ \E\left[\ \hat{u}^{ij}(\ \hat{Q}^j_{t}(V_n)\ )\ |\ \{V_n\in (\hat{Q}^k_{s})^{-1}(y)\}\ \right].
\end{equation}
Note that $D_n(T_n)=\bar{D}_n$.
The analytic expression of $D_n$ depends on the specific family of probability distribution
$\{\pi_{t}^j;t\in\tau\}$, $j\in\{1,..,d\}$, that models the relation among the response $\xi_n^j$ and the covariate profile $T_n$.
See Section~\ref{section_analyitic_expression} in the supplement \cite{AleSupp}.

Summarizing, for any $t\in\tau$ and $i\in\{1,..,d\}$, we replace in the urn identified by $t$ a number of balls of color $i$ equal to $\sum_{j=1}^dD^{ij}_n(t)X_n^j(t)$.
Hence, for any $n\geq1$ the functional urn is updated as follows:
\begin{equation}\label{eq:urn_dynamics}
\mathbf{Y}_n\ =\ \mathbf{Y}_{n-1}\ +\ D_n\mathbf{X}_n,
\end{equation}
and we set $\mathbf{Z}_{n}=\mathbf{Y}_n/\Tr(\mathbf{Y}_n)$.
Finally, we define the $\sigma$-algebra $\mathcal{F}_n$ generated
by the quantities related with the first $n$ subjects:
\begin{equation*}
\mathcal{F}_n\ :=\ \sigma\left(\mathcal{F}_{n-1},T_n,\mathbf{\bar{X}}_n,\bar{\xi}_n\right),
\end{equation*}
and we compute the probability distribution of the covariate profile of the next patient
$\mu_{n}(dt):=\P(T_{n+1}\in dt|\mathcal{F}_{n})$ with the information in $\mathcal{F}_n$.

The key feature of the design is that quantile functions are used to update \textit{all} urns, not just the urn for which
$T_n=t$.  In theory there could be an uncountably infinite number of urns, with only a finite subset of them used for patient allocation.
However, in clinical practice, mathematically ``continuous''
covariates are really not continuous \cite{RosLac}; for instance, cholesterol is represented by integer values, likely
in some range, that would, for all intents and purposes, make it a finite discrete covariate.  However, the procedure is well-defined for
uncountably infinite urns, and first order asymptotic properties can be obtained, although some of the covariate-specific metrics do not make
sense in that context.  When we move to second-order asymptotics, we partition $\tau$ into $K$ strata, which could be intervals of a continuous set.

\begin{rem}
Suppose $\pi_t^1,..,\pi_t^d$ are known, then
$D_n$ could be replaced by $D^{*}_n$, which does not depend on $\mathcal{F}_{n-1}$.
In that case, the distribution of $\mathbf{Y}_{n}$, conditionally on $\mathcal{F}_{n-1}$, depends only on $\mathbf{Y}_{n-1}$:
the functional urn composition $\{\mathbf{Y}_{n};n\geq1\}$ is a Markov process.
However, for any $t_0\in\tau$, the distribution of the real random variable $\mathbf{Y}_{n}(t_0)$, conditionally on $\mathcal{F}_{n-1}$,
depends on all the quantities contained in $\mathcal{F}_{n-1}$ given in~\eqref{def:sigma_algebra_n-1} and not only on $\mathbf{Y}_{n-1}(t_0)$.
Hence, for any $t_0\in\tau$, the real-valued sequence of the urn composition $\{\mathbf{Y}_{n}(t_0);n\geq1\}$ is not a Markov process.
This emphasizes our choice of a functional urn model.
\end{rem}

\section{Assumptions and main results}\label{section_assumptions_results}

This section is concerned with the assumptions and the main results of the design described in Section~\ref{section_model}.
Specifically, we are interested in the asymptotic behavior of the following processes:
\begin{itemize}
\item[(i)] the probability of allocation of the subjects for each covariate profile: $\{\mathbf{Z}_n(t);t\in\tau\}$;
\item[(ii)] the proportion of subjects associated with each covariate profile assigned to the treatments:
$\{\mathbf{N}_{t,n}/\Tr(\mathbf{N}_{t,n});t\in\tau\}$, where $\mathbf{N}_{t,n}:=\sum_{i=1}^n\bar{\mathbf{X}}_i\ind_{\{T_i=t\}}$;
\item[(iii)] the proportion of subjects assigned to the treatments: $\mathbf{N}_n/n$, where $\mathbf{N}_n:=\sum_{i=1}^n\bar{\mathbf{X}}_i$;
\item[(iv)] the adaptive estimators of features of interest
related with the distribution of the treatments responses conditionally on each covariate profile:
$$\{\hat{\theta}_{t,n};t\in\tau\}:=\{(\hat{\mathbf{\theta}}^j_{t,n},j\in\{1,..,d\})^\top;t\in\tau\},$$
where each estimator $\hat{\mathbf{\theta}}^j_{t,n}$ is computed with the responses of the first $N_{t,n}^j$ subjects assigned to treatment $j$
with covariate profile $t$, i.e. $\{1\leq i\leq n:\bar{X}_i^j\ind_{\{T_i=t\}}=1\}$;
\item[(v)] the adaptive estimators of features of interest
related with the entire family of distribution of the treatments responses conditionally on the covariates:
$$\hat{\beta}_{n}:=(\hat{\mathbf{\beta}}^j_{n},j\in\{1,..,d\})^\top;$$
each estimator $\hat{\mathbf{\beta}}^j_{n}$ is now computed with the responses of the first $N_{n}^j$ subjects assigned to treatment $j$,
i.e. $\{1\leq i\leq n:\bar{X}_i^j=1\}$.
\end{itemize}

\begin{rem}
In the case where $\tau$ is continuous, metrics in (ii) and (iv) have no meaning.
\end{rem}

\subsection{First-order asymptotic properties}\label{subsection_first_order_properties}

\subsubsection{Assumptions}\label{subsubsection_first_order_assumptions}
We start by providing the main assumptions that are required for establishing the first-order asymptotic properties.
\begin{assum}
\item\label{assum:1} \textbf{Constant balance and positiveness of replacement matrices.}
Let $D^{ij}_n(t)>0$ for any $i,j\in\{1,..,d\}$ and $t\in\tau$,
which is equivalent to require that $u_t^{ij}(y)>0$ for any $y\in S^j$.
Moreover, denoting $\breve{\mathbf{D}}^{\cdot j}(t)$ the $j^{th}$ column of $\breve{D}(t)$, we require that
there exists a function $c(t)$ such that $\inf_{t\in\tau}c(t)>0$ and
for any $t\in\tau$
\begin{equation}\label{ass:constant_balance}
\mathbb{P}\left(\ \Tr(\breve{\mathbf{D}}^{\cdot1}(t))=\Tr(\breve{\mathbf{D}}^{\cdot2}(t))=..=\Tr(\breve{\mathbf{D}}^{\cdot d}(t))=c(t)\ \right)\ =\ 1.
\end{equation}
\end{assum}
Since by~\eqref{def:D_breve_n_statistical_part} $\breve{D}_n^{ij}(t)=u^{ij}_t(W_t^j)$ with $W_t^j\sim \pi_t^j$,~\eqref{ass:constant_balance} holds when the function $u^{ij}_t$ is chosen such that $\sum_{i=1}^du^{ij}_t(W_t^j)$ is equal to $c(t)$
with probability one for all $j\in \{1,..,d\}$ and $t\in\tau$.
To avoid unessential complications, without loss of generality we assume throughout all the paper that $c(t)=1$.

\begin{assum}
\item\label{assum:2} \textbf{Limiting generating matrix.}
For any $t\in\tau$, let $H(t):=\E[\breve{D}_1(t)]$ and
$H_n(t):=\E[ D_n(t)| \F_{n-1},T_n,\mathbf{\bar{X}}_n]$; then, we assume that $H(t)$ is irreducible, diagonalizable and
there exists $\alpha>0$ independent of $t\in\tau$ such that
\begin{equation}\label{ass:H_minus_Hstar_finite}
\E\left[| H_n(t)-H(t) |\Big|\F_{n-1}\right]\ =\ O(n^{-\alpha})\qquad a.s.
\end{equation}
We will refer to $H_n$ as \emph{generating matrix} and to $H$ as \emph{limiting generating matrix}.
\end{assum}
A simple interpretation of~\eqref{ass:H_minus_Hstar_finite} can be obtained by noticing that $H$ can be
expressed as $\E[D^*_n|\F_{n-1},T_n,\mathbf{\bar{X}}_n]$ for all $n\geq1$,
from which it follows that the assumption in~\eqref{ass:H_minus_Hstar_finite} is related to the properties of consistency of the adaptive estimators.
To see that $H\equiv\E[D^*_n|\F_{n-1},T_n,\mathbf{\bar{X}}_n]$, observe that $\E[D^*_n|\F_{n-1},T_n,\mathbf{\bar{X}}_n]=\E[\breve{D}_n|\F_{n-1},T_n,\mathbf{\bar{X}}_n]$ and
by~\eqref{def:D_breve_n_statistical_part} $\breve{D}_n^{ij}(t)=u_t^{ij}(Q^j_{t}(V_n))$,
with $V_n\sim U(0,1)$ independent of $\mathcal{F}_{n-1}$, $T_n$ and $\mathbf{\bar{X}}_n$, which implies
$$\E\left[ \breve{D}^{ij}_n(t)| \mathcal{F}_{n-1},T_n,\mathbf{\bar{X}}_n \right]\ =\
\E\left[ u_t^{ij}( Q^j_{t}(V_n) ) \right]\ =\ H^{ij}(t).$$

\begin{rem}
Assumption~\ref{assum:2} ensures that the conditional increments of the urn composition
$\E [ \mathbf{Y}_{n}-\mathbf{Y}_{n-1} | {\F}_{n-1}]$ are
asymptotically equal to $H\mathbf{Z}_{n-1}$.
Indeed, combining~\eqref{eq:expetcted_X_equal_Z} and~\eqref{eq:urn_dynamics}, we have
\begin{equation*}
\E [ \mathbf{Y}_{n}-\mathbf{Y}_{n-1} | {\F}_{n-1}]=\E [ H_n \mathbf{X}_{n} | {\F}_{n-1}]=
O(n^{-\alpha})\ +\ H\mathbf{Z}_{n-1}.
\end{equation*}
From a probabilistic point of view, this is a key point to develop a functional urn asymptotic theory.
\end{rem}

\begin{rem}
Our functional urn model defines $H_n$ as $\E[ D_n| \F_{n-1},T_n,\mathbf{\bar{X}}_n]$
instead of $\E [ D_{n} | {\F}_{n-1}]$, typically given in the literature (e.g. see~\cite{BaiHu99,BaiHu05,LarPag}).
The two definitions coincide only when no functional dependence occurs.
However, Assumption~\ref{assum:2} ensures that the limiting generating matrix is still the same:
$$\E [ D_{n} | {\F}_{n-1}]\ =\ \E [ H_n | {\F}_{n-1}]\ =\ H\ +\ \E [ H_n-H | {\F}_{n-1}]\ \stackrel{a.s.}{\rightarrow}\ H.$$
\end{rem}

\begin{rem}
Assumption~\ref{assum:2} is verified under mild conditions on the consistency of the adaptive estimators in $\hat{u}^{ij}_t$ and $\hat{\pi}^{j}_t$.
\end{rem}

\subsubsection{First-order asymptotic results}\label{subsubsection_first_order_results}

The main consistency results concerning the design are collected in the following theorem.
\begin{theo}\label{thm:first_order_result}
Let $\mathbf{v}(t)\in \mathcal{S}$ be the right eigenvector of $H(t)$ associated with $\lambda=1$
and assume \ref{assum:1}, \ref{assum:2}.
Then,
\begin{itemize}
\item[(a)] for any probability measure $\nu$ on $\tau$, we have:
$$\int_{\tau}\|\mathbf{Z}_{n}(t)-\mathbf{v}(t)\|\nu(dt)\stackrel{a.s.}{\rightarrow}0,$$
and hence $\mathbf{Z}_{n}(t)\stackrel{a.s.}{\rightarrow}\mathbf{v}(t)$ for any $t\in\tau$;
\item[(b)] for any $t\in\tau$ such that $\sum_{i=1}^n\mu_{i-1}(\{t\})\stackrel{a.s.}{\rightarrow}\infty$,
 we have:
 $$\|\mathbf{N}_{t,n}/\Tr(\mathbf{N}_{t,n})-\mathbf{v}(t)\|\stackrel{a.s.}{\rightarrow}0;$$
\item[(c)] if there exists a probability measure $\mu$ on $\tau$ such that
$\int_{\tau}|\mu_{n}(dt)-\mu(dt)|\stackrel{a.s.}{\rightarrow}0$,
then we have that
$$\|\mathbf{N}_n/n-\int_{\tau}\mathbf{v}(t)\mu(dt)\|\stackrel{a.s.}{\rightarrow}0.$$
\end{itemize}
\end{theo}

\begin{rem}
In the special case that the covariate profiles of the subjects $\{T_n;n\geq1\}$ is a sequence of i.i.d. random variables,
we have $\mu_{i}=\mu$ for any $i\geq0$ and hence the assumptions of (b) and (c) in Theorem~\ref{thm:first_order_result}
are immediately satisfied.
\end{rem}

\begin{rem}
Conditioning on any specific covariate profile, these results are consistent with well-known asymptotic results found in the literature on urn models
(e.g. see~\cite{AthKar68,BaiHu99,BaiHu05,BaiHuZha,ZhaHuChe,ZhaHuCheCha11,zha16}),
in which the urn proportion and the allocation proportion converges a.s. to the normalized right eigenvector of the
limiting irreducible mean replacement matrix associated with $\lambda=1$.
\end{rem}

\subsection{Second-order asymptotic properties}\label{subsection_second_order_properties}

The convergence results proved in Section \ref{subsection_first_order_properties} consider a general
covariate space $\tau$.  In order to show second-order properties, we now partition $\tau$ into $K$ finite elements, which
could, for instance, be $K$ intervals of a continuous covariate space.  This partitioning induces $K$ urns used to allocate
subjects with covariate profiles in the set $\{1,...,K\}$.  In clinical trials practice, $K$ must be considerably smaller than the total sample size.

\subsubsection{Assumptions}\label{subsubsection_second_order_assumptions}

We now present further assumptions that are required for establishing the second-order asymptotic properties.
\begin{assum}
\item\label{assum:3} \textbf{Finite partition of the covariate space.}
We assume that the covariate space $\tau$ is composed by a finite number $K\in\mathbb{N}$ of distinct elements.
When $\tau$ contains infinite elements,
we can take a partition of $\tau$, i.e. $\{\tau_1,..,\tau_K\}$
such that $\cup_k\tau_k=\tau$ and $\tau_{k_1}\cap\tau_{k_2}=\emptyset$ for $k_1\neq k_2$,
and consider these sets to be the elements of $\tau$, i.e. $\tau:=\{\tau_1,..,\tau_K\}$.
To facilitate the notation, without loss of generality in the sequel we redefine $\tau=\{1,..,K\}$ and
$\mu_{n-1}(t)=\mu_{n-1}(\{t\})=\P(T_n=t|\mathcal{F}_{n-1})$ for any $t\in\tau$.

\item\label{assum:4} \textbf{Conditional response distributions.}
The analog of the null hypothesis in classical inferential statistics is given here by
assuming that the conditional response distributions $\pi_t^1,..,\pi_t^d$ are known for any $t\in\tau$.
As a direct consequence, we have that $D_n=D^{*}_n$ and $H_n=H$ with probability one for any $n\geq1$.

\item\label{assum:5} \textbf{Eigenvalues of the limiting generating matrix.}
Denoting $\lambda_H^{*}(t)$ the eigenvalue of $Sp(H(t))\setminus\{1\}$ with largest real part,
assume that $\max_{t\in\tau}{\mathcal Re}(\lambda_H^{*}(t))<1/2$.

\item\label{assum:6} \textbf{Dynamics of adaptive estimators.}
\begin{assum}
\item\label{assum:6a} \textit{(Covariate-stratification approach)}
For some $t\in\tau$ and $j\in\{1,..,d\}$, consider that there are features of interest
$\mathbf{\theta}^j_t$ related with the distribution $\pi_t^j$ of the responses
to treatment $j$ conditionally on the covariate profile $t$.
Then, we assume that the corresponding
adaptive estimator $\hat{\mathbf{\theta}}_{t,n}^j$ is strongly consistent and its dynamics can be expressed as follows:
there exists $n_0\geq1$ such that for any $n\geq n_0$
\begin{equation}\label{SAP_theta_strata}
\hat{\mathbf{\theta}}^j_{t,n}-\hat{\mathbf{\theta}}^j_{t,n-1}=-\frac{\bar{X}_n^j\ind_{\{T_n=t\}}}{N^j_{t,n}}
(f_{t,j}(\hat{\mathbf{\theta}}^j_{t,n-1})-\Delta \mathbf{M}_{t,j,n}-\mathbf{R}_{t,j,n}),
\end{equation}
where
\begin{itemize}
\item[(i)] $f_{t,j}$ is a Lipschitz continuous function such that $f_{t,j}(\mathbf{\theta}_t^j)=0$;
\item[(ii)] $\Delta \mathbf{M}_{t,j,n}\in\F_{n}$ is a martingale increment such that
$\E[\Delta \mathbf{M}_{t,j,n} |\mathcal{F}_{n-1},T_n,\bar{X}_n^j]=0$,
and it converges stably to $\Delta \mathbf{M}_{t,j}$ with kernel $K$ independent of $\F_{n-1}$:\\
$\mathcal{L}(\Delta \mathbf{M}_{t,j,n}|\mathcal{F}_{n-1},T_n=t,\bar{X}_n^j=1)\stackrel{a.s.}{\rightarrow}K(t,j)$;
\item[(iii)] $\mathbf{R}_{t,j,n}\in\F_{n}$ is such that $n\E[\|\mathbf{R}_{t,j,n}\|^2]\rightarrow0$.
\end{itemize}
Moreover, let $f_{t,j}$ be differentiable at $\mathbf{\theta}_t^j$,
denote by $\lambda^{*}_{\theta_t^j}$ the eigenvalue of $Sp(\mathcal{D}f_{t,j}(\mathbf{\theta}_t^j))$ with largest real part and
assume that $\min_{t\in\tau}{\mathcal Re}(\lambda^{*}_{\theta_t^j})>1/2$.
We also assume that for some $\delta>0$,
\begin{equation}\label{ass:martingale_theta_finite}
	\sup_{n\geq 1}\mathbb{E}\left[\left\|\Delta \mathbf{M}_{t,j,n}\right\|^{2+\delta}\,|\,\mathcal{F}_{n-1}\right]<+\infty \, a.s.,
 \end{equation}
and \begin{equation}\label{ass:martingale_theta_convergence}
\mathbb{E}\left[\Delta \mathbf{M}_{t,j,n}
(\Delta\mathbf{M}_{t,j,n})^\top\,|\,\mathcal{F}_{n-1}\right]\overset{a.s.}{\underset{n\rightarrow+\infty}{\longrightarrow}}\Gamma_{t,j},
\end{equation}
where $\Gamma_{t,j}$ is a symmetric positive matrix.

\item\label{assum:6b} \textit{(Covariate-adjusted approach)}
For some $j\in\{1,..,d\}$, consider that there are features of interest
$\mathbf{\beta}^j$ related with the entire family of distributions $\{\pi_t^j;t\in\tau\}$
of the responses to treatment $j$ conditionally on the covariates.
Then, we assume that the corresponding adaptive estimator $\hat{\mathbf{\beta}}_{n}^j$ is strongly consistent and its dynamics
can be expressed as follows:
\begin{equation}\label{SAP_theta_adjusted}
\hat{\mathbf{\beta}}^j_{n}-\hat{\mathbf{\beta}}^j_{n-1}=-\frac{\bar{X}_n^j}{N^j_n}
(f_{j}(\hat{\mathbf{\beta}}^j_{n-1})-\Delta \mathbf{M}_{j,n}-\mathbf{R}_{j,n}),
\end{equation}
where the quantities in~\eqref{SAP_theta_adjusted} fulfill the same conditions presented above for the dynamics~\eqref{SAP_theta_strata}.
\end{assum}

\begin{rem}\label{rem:popular_estimators_assumptions}
Assumption~\ref{assum:6} is usually satisfied in most relevant cases (see \cite{TouRenAir14} for the generalized urn model).
For instance, a sufficient condition for the consistency of $\hat{\theta}^j_{t,n}$, which is generally true for most practical situations, is that the symmetric part of $\mathcal{D}f_{t,j}$ is positive definite in the entire parameter space. This can be proved using analogous arguments to those used in the proof of part (a) of Theorem~\ref{thm:first_order_result}.
As an example, whenever $\theta_t^j=\E[g(\xi_n^j)|T_n=t]$ for some function $g$,
the sample mean estimator
$$\hat{\theta}^j_{t,n}=(N_{t,n}^j)^{-1}\sum_{i=1}^n\ind_{\{T_i=t\}}\bar{X}_i^jg(\bar{\xi}_i),$$
satisfies~\ref{assum:6a} with
$\Delta M_{t,j,n}=(g(\bar{\xi}_n)-\theta^j_t)$ and  $f_{t,j}(\hat{\theta}^j_{t,n-1})=(\hat{\theta}^j_{t,n-1}-\theta^j_t)$,
which implies $\mathcal{D}f_{t,j}=I$.
\end{rem}

\begin{rem}\label{rem:alternative_estimators_assumptions}
When some parameters are related to the response distribution of more than one treatment,
the corresponding estimators satisfy analogous conditions to those in~\ref{assum:6}.
Specifically, if one parameter does not depend on the treatment,
instead of~\eqref{SAP_theta_strata} we can consider
\begin{equation}\label{SAP_theta_strata_alternative}
\hat{\mathbf{\theta}}_{t,n}-\hat{\mathbf{\theta}}_{t,n-1}=-\frac{\ind_{\{T_n=t\}}}{\Tr(\mathbf{N}_{t,n})}
(f_{t}(\hat{\mathbf{\theta}}_{t,n-1})-\Delta \mathbf{M}_{t,n}-\mathbf{R}_{t,n}),
\end{equation}
while instead of~\eqref{SAP_theta_adjusted} we can consider
\begin{equation}\label{SAP_theta_adjusted_alternative}
\hat{\mathbf{\beta}}_{n}-\hat{\mathbf{\beta}}_{n-1}=-\frac{1}{n}
(f(\hat{\mathbf{\beta}}_{n-1})-\Delta \mathbf{M}_{n}-\mathbf{R}_{n}).
\end{equation}
Further dynamics may be considered when some parameters depend, for instance,
on a proper subset of the possible treatments.
\end{rem}

\item\label{assum:7} \textbf{Conditional distribution of the covariates.}
\begin{assum}
\item\label{assum:7a} \textit{(Covariate-stratification approach)}
Let $\{\mu_n;n\geq0\}$ be the sequence of probability measures on $\tau$ such that $\mu_n(t)=\P(T_{n+1}=t|\F_n)$ for any $t\in\tau$.
Assume that there exists $n_0\geq1$ such that for any $n\geq n_0$
\begin{equation}\label{ass:mu_n_second_order_strata}
\mathbf{\mu}_{n}(t)=f_{\mu,t}(\hat{\mathbf{\theta}}^j_{t,n},\mathbf{N}_{t,n}/\Tr(\mathbf{N}_{t,n})),
\end{equation}
where $\{f_{\mu,t};t\in\tau\}$ are differentiable functions, $f_{\mu,t}(\cdot)\geq\epsilon$
for some $\epsilon>0$ and $\sum_{t=1}^Kf_{\mu,t}(\cdot)=1$.
Denoting $\lambda_{\mu}^{*}$ the eigenvalue of
$Sp(\sum_{s=1}^K\mathbf{v}(s)\mathcal{D}_{N}f_{\mu,s}(\mathbf{x}_0,\mathbf{\beta})^\top)$ with
the largest real part, we assume that $\max_{t\in\tau}{\mathcal Re}(\lambda^{*}_{\mu}(t))<1/2$.

\item\label{assum:7b} \textit{(Covariate-adjusted approach)}
Let $\{\mu_n;n\geq0\}$ be the sequence of probability measures on $\tau$ such that $\mu_n(t)=\P(T_{n+1}=t|\F_n)$.
Assume that there exists $n_0\geq1$ such that for any $n\geq n_0$
\begin{equation}\label{ass:mu_n_second_order_adjusted}
\mathbf{\mu}_{n}(t)=f_{\mu,t}(\hat{\mathbf{\beta}}^j_{n},\mathbf{N}_{n}/n),
\end{equation}
where $\{f_{\mu,t};t\in\tau\}$ are differentiable functions, $f_{\mu,t}(\cdot)\geq0$
and $\sum_{t=1}^Kf_{\mu,t}(\cdot)=1$.
Moreover, we assume there exists an internal point $\mathbf{x}_0\in \mathcal{S}$ that verifies $\mathbf{x}_0=\sum_{s=1}^Kf_{\mu,s}(\mathbf{x}_0,\mathbf{\beta})\mathbf{v}(s)$.
Denoting $\lambda_{\mu}^{*}$ the eigenvalue of $Sp(\sum_{s=1}^K\mathbf{v}(s)\mathcal{D}_{N}f_{\mu,s}(\mathbf{x}_0,\mathbf{\beta})^\top)$
with the largest real part, we assume that ${\mathcal Re}(\lambda^{*}_{\mu})<1/2$.
\end{assum}
\end{assum}

\begin{rem}
The condition $f_{\mu,t}(\cdot)\geq\epsilon>0$ ensures that any covariate profile $t\in\tau$ is asymptotically observable with positive probability.
This assumption is essential to study the behavior of $\hat{\mathbf{\theta}}^j_{t,n}$ and $\mathbf{N}_{t,n}$,
while it is not necessary for $\hat{\mathbf{\beta}}^j_{n}$ and $\mathbf{N}_{n}$.
For this reason, it will be required in Theorem~\ref{thm:second_order_result_strata} but not in
Theorem~\ref{thm:second_order_result_adjusted}.
\end{rem}

\begin{rem}
In the special case that the covariate profiles of the subjects $\{T_n;n\geq1\}$ is a sequence of i.i.d. random variables,
we have $\mu_{i}=\mu$ for any $i\geq0$ and hence the above assumptions
are satisfied in a straightforward manner with $f_{\mu,t}(\cdot)=\mu(t)$ and $\mathcal{D}_{N}f_{\mu,t}=\mathcal{D}_{N}f_{\beta,t}=0$ for any $t\in\tau$.
\end{rem}

\subsubsection{Second-order asymptotic results}\label{subsubsection_second_order_results}

We first provide the convergence rate and the joint asymptotic distribution
concerning the quantities of interest in the design in the framework of
covariate-stratification response-adaptive designs.
This result is established in the following central limit theorem.
We introduce the variables independent of $\sigma(\F_n;n\geq1)$:
$T\in\tau$ with distribution $\mu(t)$,
$\bar{\mathbf{X}}\in\{0,1\}^d\in\mathcal{S}$ such that $\P(\bar{X}^j=1|T)=v^j(T)$,
$D:=\E[\breve{D}|T,\bar{\mathbf{X}},\bar{\mathbf{\xi}}]$, where the distribution of $\bar{\mathbf{\xi}}$ conditioned on $\{T=t\}$
and $\{\bar{X}^j=1\}$ is $\pi_t^j$.

\begin{theo}\label{thm:second_order_result_strata}
Define $\mathbf{W}_n:=(\mathbf{Z}_n(t),\mathbf{N}_{t,n}/\Tr(\mathbf{N}_{t,n}),\hat{\mathbf{\theta}}_{t,n},t\in\tau)^\top$,
$\mathbf{W}:=(\mathbf{v}(t),\mathbf{v}(t),\mathbf{\theta}_t,t\in\tau)^\top$ and
assume \ref{assum:1}-\ref{assum:5},\ref{assum:6a},\ref{assum:7a}.
Then,
\begin{equation}\label{eq:as_strata}
\mu_n(t)\stackrel{a.s.}{\longrightarrow}\mu(t)=f_{\mu,t}(\mathbf{v}(t),\mathbf{\theta}_t),\qquad
\mathbf{W}_n\stackrel{a.s.}{\longrightarrow}\mathbf{W},
\end{equation}
\begin{equation}\label{eq:CLT_strata}
\sqrt{n}(\mathbf{W}_n-\mathbf{W})\stackrel{\mathcal{L}}{\longrightarrow}
\mathcal{N}\left( \mathbf{0} , \Sigma \right),\qquad
\Sigma\ :=\
\int_0^{\infty}e^{u(\frac{\mathbf{I}}{2}-A)}
\Gamma e^{u(\frac{\mathbf{I}}{2}-A^\top)}du,
\end{equation}
where
\begin{equation*}
A:=
\begin{pmatrix}
A_{ZZ} & 0 & 0\\
-I & I & 0\\
0 & 0 & A_{\theta\theta}\\
\end{pmatrix},\qquad \Gamma:=
\begin{pmatrix}
\Gamma_{ZZ} & \Gamma_{ZN} & \Gamma_{Z\theta}\\
\Gamma_{ZN}^\top & \Gamma_{NN} & 0\\
\Gamma_{Z\theta}^\top & 0 & \Gamma_{\theta \theta}
\end{pmatrix},
\end{equation*}
and $A_{ZZ}$, $A_{\theta\theta}$, $\Gamma_{NN}$, $\Gamma_{\theta\theta}$  are block-diagonal matrices whose $t^{th}$ block is
\begin{itemize}
\item[(i)] $A_{ZZ}^{tt}=(I-H(t)+\mathbf{v}(t)\mathbf{1}^\top)$;
\item[(ii)] $A_{\theta\theta}^{tt}$  is a block-diagonal matrices whose $j^{th}$ block is
$[A_{\theta\theta}^{tt}]^{jj}:=\mathcal{D}f_{t,j}(\mathbf{\theta}_t^j)$;
\item[(iii)] $\Gamma_{NN}^{tt}:=\mu^{-1}(t)(diag(\mathbf{v}(t))-\mathbf{v}(t)\mathbf{v}^\top(t))$;
\item[(iv)] $\Gamma_{\theta\theta}^{tt}$ is a block-diagonal matrices whose $j^{th}$ block is\\
$[\Gamma_{\theta\theta}^{tt}]^{jj}:=(v^j(t)\mu(t))^{-1}\E[\Delta\mathbf{M}_{t,j}(\Delta\mathbf{M}_{t,j})^\top|T=t,\bar{X}^j=1]$;
\end{itemize}
and $\Gamma_{ZZ}$, $\Gamma_{ZN}$, $\Gamma_{Z\theta}$ are matrices defined as follows: for any $t_1,t_2\in\tau$
\begin{itemize}
\item[(v)] $\Gamma_{ZZ}^{t_1t_2}:=\E[D(t_1)\mathbf{g}(t_1,T,\bar{\mathbf{X}})\mathbf{g}^\top(t_2,T,\bar{\mathbf{X}})D^\top(t_2)]-\mathbf{v}(t_1)\mathbf{v}^\top(t_2)$;
\item[(vi)] $\Gamma_{ZN}^{t_1 t_2}:=H(t_1)G(t_1,t_2)diag(\mathbf{v}(t_2))-\mathbf{v}(t_1)\mathbf{v}^\top(t_2)$;
\item[(vii)] $[\Gamma_{Z\theta}^{t_1t_2}]^j:= \E[D(t_1)\mathbf{g}(t_1,t_2,\mathbf{e}_j)\Delta\mathbf{M}_{t_2,j}^\top|T=t_2,\bar{X}^j=1]$;
\end{itemize}
where $\mathbf{g}$ is a $d$-multivariate function with values in $\mathcal{S}$ defined in~\eqref{def:vector_g} in the supplement \cite{AleSupp},
and $G(t_1,t_2)$ is a matrix with columns $\{\mathbf{g}(t_1,t_2,\mathbf{e}_j);j\in\{1,..,d\}\}$.
\end{theo}

We now provide the convergence rate and the joint asymptotic distribution
of the quantities interest in the design in the framework of
covariate-adjusted response-adaptive designs.
This result is established in the following central limit theorem.
\begin{theo}\label{thm:second_order_result_adjusted}
Define $\mathbf{W}_n:=(\mathbf{Z}_n(t),t\in\tau,\mathbf{N}_{n}/n,\hat{\mathbf{\beta}}_{n})^\top$,
$\mathbf{W}:=(\mathbf{v}(t),t\in\tau,\mathbf{x}_0,\mathbf{\beta})^\top$ and
assume \ref{assum:1}-\ref{assum:5},\ref{assum:6b},\ref{assum:7b}.
Then,
\begin{equation}\label{eq:as_adjusted}
\mu_n(t)\stackrel{a.s.}{\longrightarrow}\mu(t)=f_{\mu,t}(\mathbf{x}_0,\mathbf{\beta}),\qquad
\mathbf{W}_n\stackrel{a.s.}{\longrightarrow}\mathbf{W},
\end{equation}
\begin{equation}\label{eq:CLT_adjusted}
\sqrt{n}(\mathbf{W}_n-\mathbf{W})\stackrel{\mathcal{L}}{\longrightarrow}
\mathcal{N}\left( \mathbf{0} , \Sigma \right),\qquad
\Sigma\ :=\
\int_0^{\infty}e^{u(\frac{\mathbf{I}}{2}-A)}
\Gamma e^{u(\frac{\mathbf{I}}{2}-A^\top)}du,
\end{equation}
and
\begin{equation*}
A:=
\begin{pmatrix}
A_{ZZ} & 0 & 0\\
A_{NZ} & A_{NN} & A_{N\beta}\\
0 & 0 & A_{\beta\beta}\\
\end{pmatrix},\qquad \Gamma:=
\begin{pmatrix}
\Gamma_{ZZ} & \Gamma_{ZN} & \Gamma_{Z\beta}\\
\Gamma_{ZN}^\top & \Gamma_{NN} & 0\\
\Gamma_{Z\beta}^\top & 0 & \Gamma_{\beta \beta}
\end{pmatrix},
\end{equation*}
where again and $A_{ZZ}$, $A_{\beta\beta}$, $\Gamma_{\beta\beta}$  are block-diagonal matrices whose $t^{th}$ or $j^{th}$ block is
\begin{itemize}
\item[(i)] $A_{ZZ}^{tt}=(I-H(t)+\mathbf{v}(t)\mathbf{1}^\top)$;
\item[(ii)] $A_{\beta\beta}^{jj}=\mathcal{D}f_{j}(\mathbf{\beta}^j)$;
\item[(iii)] $\Gamma_{\beta\beta}^{jj}:=(\E[v^j(T)])^{-1}\E[\Delta\mathbf{M}_{j}(\Delta\mathbf{M}_{j})^\top|\bar{X}^j=1]$;
\end{itemize}
and
\begin{itemize}
\item[(iv)] $A_{NN}:=I-\sum_{s=1}^K\mathbf{v}(s)\mathcal{D}_{N}f_{\mu,s}(\mathbf{x}_0,\beta)^\top$;
\item[(v)] $A_{N\beta}:=-\sum_{s=1}^K\mathbf{v}(s)\mathcal{D}_{\beta}f_{\mu,s}(\mathbf{x}_0,\beta)^\top$;
\item[(vi)] $\Gamma_{NN}:=diag(\E[\mathbf{v}(T)])-\E[\mathbf{v}(T)]\E[\mathbf{v}^\top(T)]$;
\end{itemize}
and $A_{NZ}$, $\Gamma_{ZZ}$, $\Gamma_{ZN}$, $\Gamma_{Z\beta}$ are matrices defined as follows: for any $t_1,t_2\in\tau$
\begin{itemize}
\item[(vii)] $A_{NZ}^{t_2}:=-\mu(t_2)I$;
\item[(viii)] $\Gamma_{ZZ}^{t_1t_2}:=\E[D(t_1)\mathbf{g}(t_1,T,\bar{\mathbf{X}})\mathbf{g}^\top(t_2,T,\bar{\mathbf{X}})D^\top(t_2)]-\mathbf{v}(t_1)\mathbf{v}^\top(t_2)$;
\item[(ix)] $\Gamma_{ZN}^{t_1}:=H(t_1)\E[G(t_1,T)diag(\mathbf{v}(T))]-\mathbf{v}(t_1)\E[\mathbf{v}^\top(T)]$;
\item[(x)] $\Gamma_{Z\beta}^{t_1j}:=\E[D(t)\mathbf{g}(t_1,T,j)\Delta\mathbf{M}_{j}^\top|\bar{X}^j=1]$.
\end{itemize}
where we recall that $\mathbf{g}$ is a $d$-multivariate function with values in $\mathcal{S}$ defined in~\eqref{def:vector_g} in the supplement \cite{AleSupp},
and $G(t_1,t_2)$ is a matrix with columns $\{\mathbf{g}(t_1,t_2,\mathbf{e}_j);j\in\{1,..,d\}\}$.
\end{theo}

\begin{rem}
We recall that Theorem~\ref{thm:second_order_result_strata} allows inferential procedures based on stratified estimators,
while Theorem~\ref{thm:second_order_result_adjusted} allows inference on covariate-adjusted regression parameters
representing the covariate-adjusted treatment effect.
\end{rem}

\begin{rem}\label{rem:alternative_estimators_results}
In the hypothesis of Remark~\ref{rem:alternative_estimators_assumptions}
when $\hat{\theta}_t$ is defined as in~\eqref{SAP_theta_strata_alternative},
Theorem~\ref{thm:second_order_result_strata} holds with (iv) and (vii) in $\Gamma$ replaced by:
\begin{itemize}
\item[(iv)] $\Gamma_{\theta\theta}^{tt}:=(\mu(t))^{-1}\E[\Delta\mathbf{M}_{t}(\Delta\mathbf{M}_{t})^\top|T=t]$;
\item[(vii)] $\Gamma_{Z\theta}^{t_1t_2}:= \E[D(t_1)\mathbf{g}(t_1,t_2,\bar{\mathbf{X}})\Delta\mathbf{M}_{t_2}^\top|T=t_2]$.
\end{itemize}
Analogously, when $\hat{\beta}$ is defined as in~\eqref{SAP_theta_strata_alternative},
Theorem~\ref{thm:second_order_result_adjusted} holds with (iii) and (x) in $\Gamma$ replaced by:
\begin{itemize}
\item[(iii)] $\Gamma_{\beta\beta}:=\E[\Delta\mathbf{M}(\Delta\mathbf{M})^\top]$;
\item[(x)] $\Gamma_{Z\beta}^{t}:=\E[D(t)\mathbf{g}(t,T,\bar{\mathbf{X}})\Delta\mathbf{M}^\top]$.
\end{itemize}
\end{rem}

\begin{rem}\label{zhangequiv}
If there are no covariates (i.e., $\tau$ is a singleton), these results reduce to the model investigated in~\cite{zha16}.  In this case, $(A-I/2)$ corresponds to $Q^{\top}$ in~\cite{zha16}, and indeed
$$A_{ZZ}=(I-H+\mathbf{v}\mathbf{1}^\top)=I/2-Q_{ZZ}^{\top},$$
$$A_{NN}=I=I/2-Q_{NN}^{\top} \mbox{ and } A_{ZN}=-I=-Q_{ZN}^{\top}.$$
We compute
$$\Gamma_{NN}=diag(\mathbf{v})-\mathbf{v}\mathbf{v}^\top=\Sigma_1,$$
$$\Gamma_{ZN}=Hdiag(\mathbf{v})-\mathbf{v}\mathbf{v}^\top=H\Sigma_1,$$
$$\Gamma_{ZZ}=\E[D\bar{\mathbf{X}}\bar{\mathbf{X}}D^\top]-\mathbf{v}\mathbf{v}^\top=H\Sigma_1H^{\top}+\Sigma_2.$$
The last equation follows, since using $H\mathbf{v}=\mathbf{v}$ and denoting $V_j$ the covariance matrix of $\mathbf{D}_{\cdot j}$, i.e. the $j^{th}$ column of $D$,
we have
$$\E[D\bar{\mathbf{X}}\bar{\mathbf{X}}D^\top]=\sum_{j=1}^dv^j\E[\mathbf{D}_{\cdot j}\mathbf{D}_{\cdot j}^{\top}]=
\sum_{j=1}^dv^j(V_j+\mathbf{H}_{\cdot j}\mathbf{H}_{\cdot j}^{\top})=\Sigma_2+Hdiag(\mathbf{v})H^{\top}$$
and
$$\mathbf{v}\mathbf{v}^\top=Hdiag(\mathbf{v})H^{\top}-H\Sigma_1H^{\top}.$$
\end{rem}

\begin{Example}\label{example_simple_variance}
Consider the inferential problem of testing the equivalence of the effects of $d=2$ treatments in presence $K=2$ covariate profiles,
under the following CDFs of the responses: $F^j_1(y)=y^{1/\alpha}$ when $T=1$ and $F^j_2(y)=y^{1/\beta}$ when $T=2$,
where $\alpha$ and $\beta$ are positive parameters.
Since $\log(1/\bar{\xi}_i)$ is an exponential random variable with mean $\alpha$ when $T=1$,
we consider the following adaptive estimator:
$$\hat{\alpha}_n=\frac{\sum_{i=1}^n\ind_{\{T_i=1\}}\log(\bar{\xi}_i^{-1})}{\sum_{i=1}^n\ind_{\{T_i=1\}}},$$
which satisfies~\eqref{SAP_theta_strata_alternative} in~\ref{assum:6} with $f_1(\hat{\alpha}_{n-1})=(\hat{\alpha}_{n-1}-\alpha)$ and
$\Delta M_{1,n}=(\log(\bar{\xi}_n^{-1})-\alpha)$. Analogous arguments hold to construct the estimator $\hat{\beta}_n$ of $\beta$.
In this case $\mathcal{D}f_1=\mathcal{D}f_2=1$, we have $A_{\theta\theta}=I$ and we need to compute $H(t)$ to find $A_{ZZ}$.
To this end note that,
since the QFs are $Q^j_1(v)=v^{\alpha}$ and $Q^j_2(v)=v^{\beta}$, we have
\[D_n(1)=\begin{pmatrix}
 V_n^{\alpha} & 1-V_n^{\alpha}\\
1-V_n^{\alpha} & V_n^{\alpha}
\end{pmatrix},\qquad
D_n(2)=\begin{pmatrix}
 V_n^{\beta} & 1-V_n^{\beta}\\
1-V_n^{\beta} & V_n^{\beta}
\end{pmatrix},\]
where we recall that $V_n\sim U(0,1)$,
which implies
\[H(1)=(1+\alpha)^{-1}\begin{pmatrix}
1 & \alpha\\
\alpha & 1
\end{pmatrix},\qquad
H(2)=(1+\beta)^{-1}\begin{pmatrix}
1 & \beta\\
\beta & 1
\end{pmatrix}.\]
From the structure of $H(t)$ above, we obtain $\mathbf{v}(t)=(1/2,1/2)^\top$ for any $t\in\{1,2\}$
and the condition $\{\max_{t}\lambda_H^{*}(t)<1/2\}$ in~\ref{assum:5} is verified for $\alpha,\beta>1/3$.
Hence, we can compute
$$A_{ZZ}^{11}=\begin{pmatrix}
1/2+\alpha & 1/2-\alpha\\
1/2-\alpha & 1/2+\alpha
\end{pmatrix},\qquad
A_{ZZ}^{22}=\begin{pmatrix}
1/2+\beta & 1/2-\beta\\
1/2-\beta & 1/2+\beta
\end{pmatrix}.$$
Since $\mathbf{v}(t)=(1/2,1/2)^\top$ implies $\mathbf{g}(t_1,t_2,\mathbf{e}_j)=\mathbf{e}_j$ for any $t_1,t_2$,
we obtain
$$\Gamma_{ZZ}^{11}=[(2\alpha+1)(\alpha+1)]^{-1}\begin{pmatrix}
2\alpha^2+\alpha+1 & 2\alpha\\
2\alpha & 2\alpha^2+\alpha+1
\end{pmatrix},$$
\[\begin{aligned}&\Gamma_{ZZ}^{12}=\ [(\alpha+\beta+1)(\alpha+1)(\beta+1)]^{-1}\times\\
&\begin{pmatrix}
\alpha^2+\alpha+\beta^2+\beta & 2(\alpha+1)(\beta+1)+(\alpha+\beta+1)(\alpha\beta+1)\\
2(\alpha+1)(\beta+1)+(\alpha+\beta+1)(\alpha\beta+1) & \alpha^2+\alpha+\beta^2+\beta
\end{pmatrix},\end{aligned}\]
while $\Gamma_{ZZ}^{22}$ is the same as $\Gamma_{ZZ}^{11}$ with $\alpha$ replaced by $\beta$.
Then, defining the $2\times2$-matrix $J:=(2I-\mathbf{1}\mathbf{1}^\top)$, we have $\Gamma_{NN}^{11}=(4\mu(1))^{-1}J$ and
$\Gamma_{NN}^{22}=(4\mu(2))^{-1}J$, while for any $t=1,2$ we have
$\Gamma_{ZN}^{1t}=(\alpha-1)/(2\alpha+2)J$ and $\Gamma_{ZN}^{2t}=(\beta-1)/(2\beta+2)J$.
Moreover, $\Gamma_{\alpha\alpha}=(\mu(1))^{-1}\alpha^2$, $\Gamma_{\beta\beta}=(\mu(2))^{-1}\beta^2$.
Finally,
$\Gamma_{Z\alpha}^{t}=\Gamma_{Z\beta}^{t}=\mathbf{0}$ for any $t\in\tau$.
\end{Example}

\section{Application to responses with Gaussian conditional distribution}\label{section_examples_linear_regression}

In this section, we analyze the functional urn model in the case that the distribution of the responses to treatments,
conditionally on the covariates, are Gaussian.
In particular, consider the following model between the covariates and the responses to treatment $j$, $j\in\{1,..,d\}$,
\begin{equation}\label{eq:linear_regression_xi}
\xi^j_n\ =\ g^j(T_n)\ +\ \epsilon^{j}_n,\ \ \ \ \forall\ n\geq1,
\end{equation}
where $g^j\in L^2(\tau)$ and $\{\epsilon^j_n;n\geq1\}\sim i.i.d.\mathcal{N}(0,\sigma_j^2)$.
We consider $g^j$ and $\sigma_j^2$ unknown and we denote by $\hat{g}^j$ and $\hat{\sigma}_j^2$ the corresponding consistent estimators.
For instance, in a parametric setting
 we may assume $g^j(t)=\sum_{i=1}^{M}\beta^j_i\phi_i(t)$ for some $M\in\mathbb{N}$, $\beta_i^j\in\mathbb{R}$ and $\phi_i\in L^2(\tau)$.
Then, letting $\sigma_j=\sigma$ for all $j\in\{1,..,d\}$, the model~\eqref{eq:linear_regression_xi} represents the classical regression analysis
with independent and homoscedastic errors.
In this case, $\hat{g}^j(t)=\sum_{i=1}^{M}\hat{\beta}^j_i\phi_i(t)$, where $\hat{\beta}_i^j$ are the least square estimators.

From~\eqref{eq:linear_regression_xi} we have that, conditionally on the set $\{T_n=t\}$,
the response $\xi^j_n$ is normally distributed with mean $g^j(t)$ and variance $\sigma_j^2$.
Hence, the family of probability distribution $\{\pi^j_{t},t\in\tau\}$ is represented by
\begin{equation}\label{eq:linear_regression_pi}
\pi^{j}_{t}\ =\ \mathcal{N}(g^j(t),\sigma_j^2)
\end{equation}
Analogously, we can define the CDF for any $y\in\mathbb{R}$ and the QF for any $v\in(0,1)$ as follows:
\begin{equation}\label{eq:linear_regression_F_Q}
F^{j}_{t}(y)\ =\ \phi\left(\ \frac{y-g^j(t)}{\sigma_j}\ \right),\ \ \ \ \ Q^{j}_{t}(v)\ =\ g^j(t)\ +\ \sigma_j z_{v},
\end{equation}
where $\phi$ and $z_v$ are, respectively, the CDF and the QF of a standard normal variable.

\subsection{Convergence to target functions}\label{subsection_linear_regression_first_order}

We now show how in the model~\eqref{eq:linear_regression_xi} the probability of assigning a patient with covariate profile $t\in\tau$, i.e. $\mathbf{Z}_n(t)$,
can converge to any desired target $\mathbf{\rho}(t)=\mathbf{\rho}(\mathbf{\theta}_t)$,
where $\mathbf{\theta}_t$ are parameters of the conditional distributions $\pi_t^j$, $j=\{1,..,d\}$.
Since $\pi^{j}_{t}$ admits a density function in $\mathbb{R}$, we can express the functional replacement matrix $D_n$ as defined in~\eqref{def:D_n_analyitic_continuous} in the supplement \cite{AleSupp}:
conditionally on $\{T_n=s\}$, $\{\bar{X}_n^k=1\}$ and $\{\xi^k_n=y\}$, we have that for any $t\in\tau$
$$D_n^{ij}(t)\ =\ \hat{u}^{ij}\left(\ \hat{g}^j(t)\ +\ \hat{\sigma}_j z_{\hat{F}^{k}_{s}(y)} \right)\ =\ \hat{u}^{ij}\left(\ \hat{g}^j(t)\ +\ \frac{\hat{\sigma}_j}{\hat{\sigma}_k}(y-\hat{g}^k(s))\ \right).$$
The consistency of the estimators ensures that Assumption~\ref{assum:2} is satisfied and
the asymptotic behavior of the urn process is determined by the limiting generating matrix $H$ defined as
$$ H^{ij}(t)\ =\ \E\left[\ \breve{D}^{ij}_n(t)\ \right]\ =\ \E\left[\ u^{ij}(\ Q^j_{t}(V_n)\ )\ \right]\ =\ \E\left[\ u^{ij}(\ g^j(t)\ +\ \sigma_j z_{V_n}\ )\ \right],$$
where $z_{V_n}\sim\mathcal{N}(0,1)$ since $V_n\sim U(0,1)$.
Thus, from Theorem~\ref{thm:first_order_result} we have that $\mathbf{Z}_n(t)\stackrel{a.s.}{\rightarrow}\mathbf{v}(t)$, where $\mathbf{v}$ is such that $\Tr(\mathbf{v})=1$ and $H\mathbf{v}=\mathbf{v}$.
Hence, the functions $u^{ij}$, $i,j\in\{1,..,d\}$, can be chosen such that $\mathbf{v}(t)$ coincides
with the desired target function $(\rho^1(t),..,\rho^d(t))^\top$.

In the case when $d=2$ treatments, we now consider the target proportion allocation proposed in~\cite{ZhangRos}
for responses distributed as $\mathcal{N}(m_1,\sigma_1)$ for treatment 1 and $\mathcal{N}(m_2,\sigma_2)$ for treatment 2:
\begin{equation}\label{eq:target_allocation_zhangros}
\tilde{\rho}(m_1,m_2,\sigma_1,\sigma_2)\ =\ \frac{\sigma_1\sqrt{m_2}}{\sigma_1\sqrt{m_2}+\sigma_2\sqrt{m_1}},\ \ \ \ \ m_1,m_2>0.
\end{equation}
As described in~\cite{ZhangRos}, the allocation proportion~\eqref{eq:target_allocation_zhangros} minimizes
the total expected responses from all the subjects ($n\left(m_1\cdot \tilde{\rho}+m_2\cdot (1-\tilde{\rho})\right)$) with a fixed variance,
($n^{-1}\left(\sigma_1^2/\tilde{\rho} + \sigma_2^2/(1-\tilde{\rho})\right)$).
In our framework, the target proportion is the function $(\rho(t),1-\rho(t))^\top$, where $\rho(t)=\tilde{\rho}(g^1(t),g^2(t),\sigma_1,\sigma_2)$.
To achieve this limiting proportion we need to define the functions $u_t^{11}, u_t^{12}, u_t^{21}$ and $u_t^{22}$
such that the normalized right eigenvector of $H(t)$ is $\mathbf{v}(t)=(\rho(t),1-\rho(t))^\top$.
For instance, a possible choice is the following: for any $y\in\mathbb{R}$,
$$\hat{u}^{11}_t(y)\ =\ \tilde{\rho}\left(\ \hat{g}^1(t)\ ,\ \hat{g}^2(t)\ ,\ \hat{\sigma}_1\ ,\ \hat{\sigma}_2\ \right),$$
$\hat{u}^{12}_t=\hat{u}^{11}_t$ and $\hat{u}^{21}_t=\hat{u}^{22}_t=1-\hat{u}^{11}_t$.
Hence, we have that $H^{11}(t)=\lim_{n\rightarrow\infty}\E[\hat{u}^{11}_t(\xi_n^1)|T_n=t]=\rho(t)$ a.s. and,
analogously, $H^{12}(t)=\rho(t)$ and $H^{21}(t)=H^{22}(t)=1-\rho(t)$, which implies $\mathbf{v}=(\rho(t),1-\rho(t))^\top$.

\subsection{Inference on conditional response distribution}\label{subsection_linear_regression_second_order}

We now analyze how to do inference in the model~\eqref{eq:linear_regression_xi}.
Specifically, we consider the problem of testing the equivalence of the response means conditionally on the covariates,
i.e. $H_0:g^j=g$ for any $j\in\{1,..,d\}$, with $g$ given function in $L^2(\tau)$.
Take $d=2$ treatments, consider $K\geq1$ possible covariate profiles and assume $\sigma_1=\sigma_2=\sigma\in(0,\infty)$ to be known.
We set
$$u^{11}_t(y)\ =\ u^{22}_t(y)\ =\ \phi\left(\frac{y-g(t)}{\sigma}\right),$$
$u^{21}_t=1-u^{11}_t$ and $u^{12}_t=1-u^{22}_t$.
From this choice, under $H_0$ we have that: $D^{jj}_n(t)=\phi((\xi^j_n-g(t))/\sigma)\sim U(0,1)$ and for $i\neq j$
$D^{ij}_n=1-D^{jj}_n\sim U(0,1)$,
which implies $H^{ij}(t)=1/2$ for any $i,j\in\{1,2\}$ and hence $\mathbf{v}(t)=(1/2,1/2)^\top$ and $\lambda_H^{*}(t)=0$ for any $t\in\tau$.
Then, we can apply the CLT established in Theorem~\ref{thm:second_order_result_adjusted}
to construct inferential procedures to test the null hypothesis.
It is worth seeing how the dynamics of the functional urn model changes when $H_0$ does not hold.
In particular, under $H_1:\{g^1=g+\Delta\}$, for some $\Delta\in L^2(\tau)$, we have
$$D^{11}_n(t)\ =\ \phi\left(\frac{\xi^1_n-g^1(t)}{\sigma}+\frac{\Delta(t)}{\sigma}\right)\ \sim\ \phi\left(z_{V_n}+\frac{\Delta(t)}{\sigma}\right),$$
where $z_{V_n}\sim \mathcal{N}(0,1)$.

\section{Application to responses with Bernoulli conditional distribution}\label{section_examples_logistic_regression}

In this section, we analyze the functional urn model when the responses to treatments,
conditionally on the covariates, are Bernoulli distributed.
In particular, consider the following model between the covariates and the responses to treatment $j$, $j\in\{1,..,d\}$,
\begin{equation}\label{eq:logistic_regression_xi}
\xi^j_n\ =\ \ind_{\{U^j_n\leq p^j(T_n)\}},\ \ \ \ \forall\ n\geq1,
\end{equation}
where $U^j_n\ \sim\ i.i.d.\ U(0,1)$ and $0< p^j(t)< 1$ for any $t\in\tau$.
We consider $p^j$ unknown and we denote by $\hat{p}^j$ its consistent estimator.
If we assume there exist $M\in\mathbb{M}$, $\beta_i^j\in\mathbb{R}$ and $\phi_i\in L^2(\tau)$ such that
\begin{equation}\label{eq:logistic_regression_p}
\log\left(\frac{p^j(t)}{1-p^j(t)}\right)\ =\ \sum_{i=1}^{M}\beta^j_i\phi_i(t),
\end{equation}
the model~\eqref{eq:logistic_regression_xi} represents the classical logistic regression for binary responses.
In this case, $\hat{p}^j(t)=\sum_{i=1}^{M}\hat{\beta}^j_i\phi_i(t)$, where $\hat{\beta}_i^j$ are the maximum likelihood estimators (MLEs) of $\beta_i^j$.

From~\eqref{eq:logistic_regression_xi} we have that, conditionally on the set $\{T_n=t\}$,
$\{\xi^j_n;n\geq1\}$ represents a sequence of independent Bernoulli random variables with parameter $p^j(t)$.
Hence, the probability measures $\{\pi^j_{t},t\in\tau\}$ are Bernoulli distributed as
\begin{equation}\label{eq:logistic_regression_pi}
\pi^{j}_{t}\ =\ \mathcal{B}e(\ p^j(t)\ ),\ \ \ \ \forall\ n\geq1.
\end{equation}
Analogously, we can define the CDF for any $y\in\mathbb{R}$ and the QF for any $v\in(0,1)$
\begin{equation}\label{eq:logistic_regression_F_Q}
F^{j}_{t}(y)\ =\ (1-p^j(t))\ind_{\{y\geq0\}}\ +\ p^j(t)\ind_{\{y\geq1\}},\ \ \ \ \ Q^{j}_{t}(v)\ =\ \ind_{\{v\geq1- p^j(t)\}}.
\end{equation}

\subsection{Convergence to target functions}\label{subsection_logistic_regression_first_order}

Since $\pi^{j}_{t}$ is a discrete distribution, the functional replacement matrix $D_n$ can be expressed as in~\eqref{def:D_n_analyitic_discrete} in the supplement \cite{AleSupp}:
conditionally on $\{T_n=s\}$, $\{\bar{X}_n^k=1\}$ and $\{\xi^k_n=y\}$, for any $t\in\tau$ we have that
\[\begin{aligned}
D_n^{ij}(t) & =
\begin{cases}
(1-\hat{p}^k(s))^{-1}\cdot\int_{0}^{1-\hat{p}^k(s)}\ \hat{u}^{ij}(\ind_{\{v\geq1- \hat{p}^j(t)\}})dv\ & \text{if } y=0;\\
(\hat{p}^k(s))^{-1}\cdot\int_{1-\hat{p}^k(s)}^1\ \hat{u}^{ij}(\ind_{\{v\geq 1-\hat{p}^j(t)\}})dv\ & \text{if } y=1.
\end{cases}
\end{aligned}\]
The consistency of the estimators ensures that Assumption~\ref{assum:2} is satisfied and
the asymptotic behavior of the urn process is determined by the limiting generating matrix $H$ defined as
$$ H^{ij}(t)\ =\ \E\left[\ \breve{D}^{ij}_n\ \right]\ =\ \E\left[\ u^{ij}(\ Q^j_{t}(V_n)\ )\ \right]\ =\ \E\left[\ u^{ij}(\ \ind_{\{V_n\geq1- p^j(t)\}}\ )\ \right],$$
where $V_n\sim U(0,1)$.
Thus, from Theorem~\ref{thm:first_order_result} we have that $\mathbf{Z}_n(t)\stackrel{a.s.}{\rightarrow}\mathbf{v}(t)$, where $\mathbf{v}$ is such that $\Tr(\mathbf{v})=1$ and $H\mathbf{v}=\mathbf{v}$.
Hence, the functions $u_t^{ij}$, $i,j\in\{1,..,d\}$, can be chosen such that $\mathbf{v}(t)$ coincides with the desired target function
$(\rho^1(t),..,\rho^d(t))^\top$.

For instance, consider play-the-winner design for binary responses proposed in~\cite{WeiDur} and~\cite{Zel69}.
In the multi-treatments play-the-winner design, when treatment $j\in\{1,..,d\}$ is assigned,
we replace in the urn a ball of color $j$ if the response is a success or $(d-1)^{-1}$ balls of each other color if the response is a failure.
Thus, the play-the-winner rule can be implemented in our framework by setting $u_t^{ij}(y)=y\delta_{ij}+(1-y)(1-\delta_{ij})(d-1)^{-1}$,
for any $y\in\{0;1\}$, where $\delta_{ij}$ is the delta of kronecker.
Note that this choice of $u_t^{ij}$ guarantees the constant balance of the urn required in~\ref{assum:1}.
Then, each element $D_n^{ij}(t)$ of the replacement matrix can be explicitly expressed as follows:
\[\begin{cases}
\left(\frac{\max\{\hat{p}^j(t);\hat{p}^k(s)\}-\hat{p}^k(s)}{1-\hat{p}^k(s)}\right)\delta_{ij}
+\left(\frac{1-\max\{\hat{p}^j(t);\hat{p}_k(s)\}}{1-\hat{p}^k(s)}\right)(1-\delta_{ij})(d-1)^{-1} & \text{if } y=0;\\
\left(\frac{\min\{\hat{p}^j(t);\hat{p}_k(s)\}}{\hat{p}^k(s)}\right)\delta_{ij}+\left(\frac{\hat{p}^k(s)
-\min\{\hat{p}^j(t);\hat{p}_k(s)\}}{\hat{p}^k(s)}\right)(1-\delta_{ij})(d-1)^{-1} & \text{if } y=1.
\end{cases}\]
In this case, since $\ind_{\{V_n\geq1- p^j(t)\}}\sim \mathcal{B}e(p^j(t))$ when $V_n\sim U(0,1)$, we have
\[\begin{aligned}
H^{ij}_{n-1}(t)\ &&=&\ \E\left[\ \ind_{\{V_n\geq1- p^j(t)\}}\delta_{ij}\ +\ (1-\ind_{\{V_n\geq1- p^j(t)\}})(1-\delta_{ij})(d-1)^{-1}\ \right]\\
&&=&\ p^j(t)\delta_{ij}\ +\ (1-p^j(t))(1-\delta_{ij})(d-1)^{-1}.
\end{aligned}\]
Hence, the first right eigenvector $\mathbf{v}(t)$ of $H(t)$ associated with the eigenvalue $\lambda=1$ is
\begin{equation}\label{eq:target_allocation_play_the_winner}
v^{j}(t)\ =\ \rho^j(p^1(t),..,p^d(t))\ = \frac{(1-p^j(t))^{-1}}{\sum_{i=1}^{d}(1-p^i(t))^{-1}},
\end{equation}
and from Theorem~\ref{thm:first_order_result} we have that $\mathbf{Z}_n(t)\stackrel{a.s.}{\rightarrow}(\rho^1(t),..,\rho^d(t))^\top$.

\subsection{Inference on conditional response distribution}\label{subsection_logistic_regression_second_order}

We now analyze how to do inference in the model~\eqref{eq:logistic_regression_xi}.
Specifically, we consider the problem of testing the equivalence of the success probabilities conditionally on the covariates,
i.e. $H_0:p^j=p$ for any $j\in\{1,..,d\}$, with $p$ given function with values in $(0,1)$.
Using the same functions defined above, i.e. $u_t^{ij}(y)=y\delta_{ij}+(1-y)(1-\delta_{ij})(d-1)^{-1}$,
we have that, under $H_0$,
$$D^{ij}(t)  = \delta_{ij}\ind_{\{\bar{\xi}_n=1\}} + (1-\delta_{ij})\ind_{\{\bar{\xi}_n=0\}}(d-1)^{-1}\ \sim\
\delta_{ij}W + (1-\delta_{ij})(1-W)(d-1)^{-1},$$
where $W\sim \mathcal{B}e(p)$.
This implies $H^{jj}(t)=p(t)$ and $H^{ij}(t)=(1-p(t))(d-1)^{-1}$ for $i\neq j$, and hence
$\mathbf{v}(t)=d^{-1}\mathbf{1}$ and $\lambda_H^{*}(t)=(dp(t)-1)/(d-1)$ for any $t\in\tau$.
Then, if $\max_{t\in\tau}p(t)<(d+1)/(2d)$ we have $\max_{t\in\tau}\lambda_H^{*}(t)<1/2$ and hence
we can apply the CLT established in Theorem~\ref{thm:second_order_result_adjusted}
to construct inferential procedures to test the null hypothesis.
It is worth seeing how the dynamics of the functional urn model changes when $H_0$ does not hold.
In particular, under $H_1:\{p^1=p+\Delta\}$, for some $\Delta$ with values in $(0,1)$, we have
we have $D^{i1}(t)\sim\delta_{i1}W + (1-\delta_{i1})(1-W)(d-1)^{-1}$, with $W\sim \mathcal{B}e(p+\Delta)$.

\subsection{Example}\label{subsection sepsicool}

We provide a very simple example from a clinical trial of external cooling in patients with septic shock (the Sepsicool trial \cite{sepsicool}).
The trial found little difference in the primary outcome in the entire clinical trials population, but 14 day mortality, a different endpoint,
was significantly lower among patients given external cooling in the subgroup with a lower baseline vasopressor dose (i.e., those patients who had less severe
illness at baseline).  We take the approach of \cite{villar} by redesigning the study using our methodology.  Using their parameter values, obtained
from the results presented in \cite{sepsicool}, we determine that the underlying probability of survival in the ``no cooling'' group is 0.657 regardless
of severity; the probability of survival in the cooling group is 0.842 for those with low severity, and 0.406 for those with high or moderate severity.  As in
\cite{villar}, we assume that 225 patients have low severity and 225 patients have high or moderate severity.  We now redesign the trial using our methodology.
For the underlying parameter set $\boldsymbol{p}=(p^1(1),p^1(2),p^2(1),p^2(2))$
is given by $(p^1(1)=p^1(2)=0.657, p^2(1)=0.842, p^2(2)=0.406)$.
Computing the asymptotic target allocation using
\eqref{eq:target_allocation_play_the_winner}, we can compute the expected number of deaths in 450 patients using our methodology is 146.5.
If we ignore the binary covariate in the urn process, the resulting expected number of deaths is 161.4.

We can conduct hypothesis testing on $p^j(t)$ using Theorem \ref{thm:second_order_result_strata}.
For the simple case of two treatments and a single binary covariate, the joint asymptotic distribution of observed proportions $\hat{\boldsymbol{p}}=(\hat{p}^1(1),\hat{p}^1(2),\hat{p}^2(1),\hat{p}^2(2))$ can be expressed as:
\begin{equation*}
\sqrt{n}(\hat{\boldsymbol{p}}_n-\boldsymbol{p})\stackrel{\mathcal{L}}{\longrightarrow}
\mathcal{N}\left( \mathbf{0} , diag(d^1(1),d^1(2),d^2(1),d^2(2)) \right),
\end{equation*}
\begin{equation*}
\mbox{where}\qquad d^j(t)\ =\ \frac{2}{v^j(t)}p^j(t)(1-p^j(t)).
\end{equation*}
Note that, when there are no covariates, this is the same asymptotic result from the generalized P\'{o}lya urn model
\cite{RosFlour}.

\begin{rem}
Several recent papers on CARA designs show that results on parameter estimation in the binary case are analogous
to those presented in Example~\ref{subsection sepsicool},
either for the parameters of the logistic regression $\beta^j$ (see e.g.~\cite{CheungZhangHuChan,ZhaHuCheCha07,Zhu}) and for the parameters of the success probability $p^j(t)$ (see e.g.~\cite{HuZhuHu}). The connection between these two approaches is also highlighted in~\cite[Section 2.2]{HuZhuHu}.
Naturally, the same results presented in Example~\ref{subsection sepsicool} can be obtained with the unified family of CARA designs recently proposed in~\cite{HuZhuHu}. Indeed, this follows by combining~\cite[Theorem 1]{HuZhuHu} and~\cite[Example 2]{HuZhuHu}, setting $e_k(\cdot)=(1-p_k(\cdot))^{-1}$ and $\gamma=0$.
\end{rem}

\section{Conclusions}\label{section_conclusions}

This paper proposes a general class of CARA designs that
randomly assigns subjects to the treatment groups with a probability that depends on
their own covariate profiles and on the previous patients' covariates, treatment assignments and responses.
This procedure can be considered as a general methodology
to incorporate both the responses and the covariates in the randomized treatment allocation scheme,
as a possible alternative to the class of designs presented in~\cite{ZhaHuCheCha07}.
The generality of the proposed framework includes several different ways
to model the relationship between covariates and treatment responses,
allowing and facilitating the implementation of these designs to a wide range of applications.
In particular, in the paper we have discussed the properties
with generalized linear models for different types of responses and covariates (discrete and continuous)
and for more than two treatments.
Moreover, this class of designs does not require that the probability distribution of the covariates be the same for all patients,
which is a standard assumption in CARA designs (e.g. see~\cite{ZhaHuCheCha07}).
In fact, we allow this distribution to be adaptively modified by the experimenter
using the information collected during the trial.
This improvement opens the possibility to apply the theory of optimal designs (see~\cite{BaldiZag}) within the CARA framework for future research.

The class of CARA designs presented here is based on a new functional urn model
that extends the classical theory of urn models adopted for responses-adaptive designs,
in which the covariate information is not considered in the randomization process (e.g. \cite{DurFloLi98,MayFlo09,ZhaHuChe,HuRos}).
The urn is represented by a multivariate function of the covariates,
and each patient is assigned by sampling from the urn evaluated at his own covariate profile.
After any allocation, the entire functional urn composition is updated,
even if only the response associated with the patient's covariate profile has been observed,
and this allows the incorporation of general covariate spaces in the design.
In the context of personalized medicine, this feature can allow the investigation to determine optimal treatments
based on covariate model, even when there is insufficient information about a particular covariate profile.

In response-adaptive randomization,
desirable ethical or inferential properties of the design are achieved by targeting
an optimal allocation proportion, which is typically determined by some optimality criterion based on the response distribution.
Analogously, when covariate information is considered in the trial,
the purpose becomes to target an optimal allocation for each fixed value of the covariates.
In this paper we achieve this goal by establishing, in Theorem~\ref{thm:first_order_result},
the convergence to any optimal allocation.
This result is obtained by allowing nonparametric or semi-parametric estimates
of the response distribution conditioned on the covariates.
This extends the class of designs proposed in~\cite{ZhaHuCheCha07}
in which the target allocation proportion depends on a finite number of parameters.

In addition, statistical inference on the treatment effects requires the establishment of the joint distribution of sufficient statistics that,
in an adaptive setting, are represented by both allocation proportion and adaptive estimators.
This is typically a hard task in the framework of CARA designs (see~\cite{RosSve}).
Theorem~\ref{thm:second_order_result_strata} and Theorem~\ref{thm:second_order_result_adjusted}
provide the theoretical results which allow us to construct inferential procedures based on two different approaches:
stratified and covariate-adjusted estimators.
The study of their power under different types of alternative hypotheses is essential to investigate the performances of these procedures
and to conduct comparisons with other existing CARA designs.
In this paper, we have provided the general framework for investigations of this type under different models.

\begin{supplement}[id=supp]
 \sname{Online supplementary materials}
 \stitle{Nonparametric covariate-adjusted response-adaptive design based on a functional urn model}
 \slink[doi]{COMPLETED BY TYPESETTER}
 \sdatatype{.pdf}
 \sdescription{This supplement gives the analytic expressions used in the paper and the proofs of the theorems.}
\end{supplement}


\setcounter{page}{1}
\setcounter{section}{0}
\renewcommand \thesection {\Alph{section}}

\title{Online Supplemental Appendix:  Nonparametric covariate-adjusted response-adaptive design based on a functional urn model}

\section{Analytic expressions}\label{section_analyitic_expression}

We now derive some useful analytic expressions for $\mathbf{X}_n$ and $D_n$.
Using~\eqref{def:X_n_statistical_part}, we can express $\mathbf{X}_n$ as follows:
on the set $\{\bar{X}_n^k=1\}$, $k\in\{1,..,d\}$, we have, for any $j\in\{1,..,d\}$,
\begin{equation}\label{def:X_n_statistical_part_simply}
\begin{aligned}
X_n^j(t)&=\P\left(\sum_{i=1}^{j-1}Z_{n-1}^{i}(t)<U_n\leq\sum_{i=1}^{j}Z_{n-1}^i(t)\ \big|\ \mathcal{F}_{n-1},\bar{X}_n^k=1\right)
\\
&
= \frac{
\left( \min\left\{\sum_{i=1}^{j}Z_{n-1}^{i}(t);\sum_{i=1}^{k}Z_{n-1}^{i}(T_n)\right\} -
\max\left\{\sum_{i=1}^{j-1}Z_{n-1}^{i}(t);\sum_{i=1}^{k-1}Z_{n-1}^{i}(T_n)\right\} \right)^+}{Z_{n-1}^{k}(T_n)},
\end{aligned}
\end{equation}
where by convention $\sum_{i=1}^{0}(\cdot)=0$.
Note that $\Tr(\mathbf{X}_n(t))=1$ for all $t\in\tau$ and $\mathbf{X}_n(T_n)=\mathbf{\bar{X}}_n$.

The definition of $D_n$ in~\eqref{def:D_n_statistical_part_true} may be simplified
when $\hat{\pi}^k_{s}$, for some $s\in\tau$ and $k\in\{1,..,d\}$, is absolutely continuous or discrete.
In fact, in the first case the QF is bijective,
i.e. $(\hat{Q}^k_{s})^{-1}(y)\equiv \hat{F}^k_{s}(y)$ for any $y\in S^k$, and hence from~\eqref{def:D_n_statistical_part_true},
on the sets $\{T_n=s\}$, $\{\bar{X}_n^k=1\}$ and $\{\xi^{k}_n=y\}$,
$D_n$ reduces to
\begin{equation}\label{def:D_n_analyitic_continuous}
D_n^{ij}(t)\ =\ \hat{u}^{ij}(\ \hat{Q}^j_{t}(\ \hat{F}^k_{s}(y)\ )\ ).
\end{equation}
When $\hat{\pi}^k_{s}$ is discrete, for some $y\in S^k$ we have $\hat{\pi}^k_{s}(y)>0$, and hence
$$(\hat{Q}^k_{s})^{-1}(y)\ =\ \left(\ \hat{F}^k_{s}(y^{-})\ ,\ \hat{F}^k_{s}(y)\ \right),$$
where $y^{-}:=(y-\epsilon)$ with $\epsilon>0$ arbitrary small.
Thus, from~\eqref{def:D_n_statistical_part_simply}, on the sets $\{T_n=s\}$, $\{\bar{X}_n^k=1\}$ and $\{\xi^{k}_n=y\}$,
$D_n$ reduces to
\begin{equation}\label{def:D_n_analyitic_discrete}
\begin{aligned}
D_n^{ij}\ &&=&\ \left(\hat{F}^k_{s}(y)-\hat{F}^k_{s}(y^{-})\right)^{-1}\int_{\hat{F}^k_{s}(y^{-})}^{\hat{F}^k_{s}(y)}\hat{u}_t^{ik}(\hat{Q}^j_{t}(v))dv\\
&&=&\ \left(\hat{\pi}^k_{s}(y)\right)^{-1}\int_{\hat{F}^k_{s}(y^{-})}^{\hat{F}^k_{s}(y)}\hat{u}_t^{ik}(\hat{Q}^j_{t}(v))dv,
\end{aligned}
\end{equation}
where we recall that $\hat{\pi}^k_{s}$ is the estimator of $\pi^k_{s}(y)=\P(\xi_{n}^k=y|T_n=s)$.

\section{Proofs}\label{section_proofs}

This section is concerned with the proofs of the results presented in Section~\ref{section_assumptions_results}.

\subsection{Proof of the first-order asymptotic results}\label{subsection_proofs_first_order_properties}


We now prove Theorem~\ref{thm:first_order_result}.
We first need to introduce some notation concerning the eigen-structure of $H(t)$.

For any $t\in\tau$, $H(t)$ is diagonalizable by~\ref{assum:2}.
Then there exists a nonsingular matrix
$\widetilde{U}(t)$ such that $\widetilde{U}^\top(t)H(t)(\widetilde{U}^\top(t))^{-1}$ is diagonal with elements $\lambda_j(t)\in Sp(H(t))$.
Notice that each column $\mathbf{u}_j(t)$ of $\widetilde{U}(t)$ is a left eigenvector of $H(t)$ associated with $\lambda_j(t)$.
WLOG, we set $\|\mathbf{u}_j\|(t)=1$.
Moreover, when the multiplicity of some $\lambda_j(t)$ exceeds one,
we assume the corresponding eigenvectors to be orthogonal.
Then if we define $\widetilde{V}(t)=(\widetilde{U}^\top(t))^{-1}$, each column $\mathbf{v}_j(t)$ of $\widetilde{V}(t)$
is a right eigenvector of $H(t)$ associated with $\lambda_j(t)$ such that
\begin{equation}\label{eq:relazioni-0}
\mathbf{u}_j^\top\,\mathbf{v}_j=1\quad\mbox{ and }\qquad
\mathbf{u}_h^\top\,\mathbf{v}_j=0,\ \forall h\neq j.
\end{equation}
These constraints, combined with the assumptions in~\ref{assum:2} on $H$
(precisely, nonnegativity, constant balance and irreducibility)
imply, by the Frobenius-Perron Theorem, that, for any $t\in\tau$, $\lambda_1(t)=1$ is an eigenvalue of $H(t)$ with multiplicity one,
$\max_{j>1}\Re e(\lambda_j)<1$ and
\begin{equation*}
\mathbf{u}_1=N^{-1/2}\mathbf{1}, \qquad N^{-1/2}{\mathbf 1}^\top{\mathbf v}_1=1,\qquad
\qquad v^j_1>0\; \forall j=1,..,d.
\end{equation*}
Because $\mathbf{v}(t)\in\mathcal{S}$, or equivalently $\Tr(\mathbf{v}(t))=1$, in the statement of Theorem~\ref{thm:first_order_result},
then $\mathbf{v}=N^{-1/2}\mathbf{v}_1$.

In the sequel, we will use $U$ and $V$ to indicate
the sub-matrices of $\widetilde{U}$ and $\widetilde{V}$, respectively,
whose columns for any $t\in\tau$ are the left and the right eigenvectors of
$H(t)$ associated with $Sp(H(t))\setminus \{1\}$, given by
$\{\mathbf{u}_2(t),..,\mathbf{u}_N(t)\}$ and
$\{\mathbf{v}_2(t),..,\mathbf{v}_N(t)\}$, respectively.

Now, given the eigen-structure of $H$ presented here,
the matrix ${\mathbf v}\mathbf{1}^\top$ has real entries and the following relations hold:
\begin{equation}\label{eq:relations}
V^\top\,\mathbf{1}=U^\top\,\mathbf{v}=\mathbf{0},\quad
V^\top U=U^\top V=I\quad\mbox{and}\quad
I={\mathbf v}{\mathbf 1}^\top + VU^\top,
\end{equation}
where the identity matrices above have dimensions $(d-1)$ and $d$, respectively.
As a consequence of~\eqref{eq:relations}, the matrix $U(t)V^\top(t)$ has real entries for any $t\in\tau$.
Moreover, denoting by $\Lambda(t)$ the diagonal matrix whose
elements are $\lambda_j(t)\in Sp(H(t))\setminus\{1\}$, we can decompose the functional
matrix $H$ as follows:
\begin{equation}\label{eq:decomposition_H_star}
H\ =\ {\mathbf v}{\mathbf 1}^\top\ +\ V\Lambda U^\top.
\end{equation}
With this notation in mind, we are now ready to present the proof of the first-order results.

\proof[Proof of Theorem~\ref{thm:first_order_result}]
The structure of the proof of part (a) is analogous to that in {\cite[Theorem 4.1]{AleCriGhi16}}.
Consider the urn dynamics expressed in~\eqref{eq:urn_dynamics} as follows:
let $\mathbf{Y}_0=\mathbf{1}$ and for any $n\geq1$
\begin{equation}\label{eq:urn_dynamics_proofs}
\mathbf{Y}_n\ =\ \mathbf{Y}_{n-1}\ +\ D_n\mathbf{X}_n.
\end{equation}
From~\eqref{eq:urn_dynamics_proofs}, we can derive the following decomposition:
\begin{equation}\label{eq:urn_dynamics_proofs_1}
\mathbf{Y}_n\ =\ \mathbf{Y}_{n-1}\ +\ H\mathbf{Z}_{n-1}\ +\ \Delta \mathbf{M}_{Z,n}\ +\ \mathbf{R}_{Z,n},
\end{equation}
where
\begin{itemize}
\item[(1)] $\Delta \mathbf{M}_{Z,n}:=(D_n-H_n)\mathbf{X}_n+H(\mathbf{X}_n-\mathbf{Z}_{n-1})$ is a martingale increment, since
$$\E[(D_n-H_n)|\F_{n-1},T_n,\mathbf{\bar{X}}_n]\ =\ \E[(\mathbf{X}_n-\mathbf{Z}_{n-1})|\F_{n-1},T_n]\ =\ 0.$$
\item[(2)] $\mathbf{R}_{Z,n}:=(H_n-H)\mathbf{X}_n$ is a remainder term that converges to zero a.s.
due to the fact that, since $\mathbf{X}_n\in\mathcal{S}$, $\P(\|\mathbf{X}_n\|\leq1)=1$ a.s.
and by Assumption~\ref{assum:2}.
\end{itemize}
Let $r_n:=(d+n)^{-1}$. By Assumption~\ref{assum:1}, $\Tr(\mathbf{Y}_{n})=(d+n)$
with probability one for any $n\geq0$,
and hence $\mathbf{Z}_{n}=r_n\mathbf{Y}_{n}$.
Then, multiplying the dynamics~\eqref{eq:urn_dynamics_proofs_1} by $r_n$ and using $r_nr^{-1}_{n-1}=(1-r_n)$,
we obtain
\begin{equation*}
\mathbf{Z}_n\ =\ [I-r_n(I-H)]\mathbf{Z}_{n-1}\ +\ r_n\Delta \mathbf{M}_{Z,n}\ +\ r_n\mathbf{R}_{Z,n}.
\end{equation*}
Moreover, since $(I-H)\mathbf{v}=\mathbf{0}$ and defining $\mathbf{W}_n:=(\mathbf{Z}_n-\mathbf{v})$,
we obtain the following expression:
\begin{equation}\label{eq:urn_dynamics_proofs_2}
\mathbf{W}_n\ =\ [I-r_n(I-H)]\mathbf{W}_{n-1}\ +\ r_n\Delta \mathbf{M}_{Z,n}\ +\ r_n\mathbf{R}_{Z,n}.
\end{equation}
Let us consider the $(d-1)$-dimensional complex process $\{\mathbf{W}_{U,n};n\geq1\}$ defined as $\mathbf{W}_{U,n}=U^\top\mathbf{W}_{n}$.
The relation $\mathbf{W}_{n}=V\mathbf{W}_{U,n}$ is a consequence of~\eqref{eq:relations} and
$\mathbf{1}^\top\mathbf{W}_{n}=(\mathbf{1}^\top\mathbf{\mathbf{Z}_n})-(\mathbf{1}^\top\mathbf{\mathbf{v}})=0$.
Hence, to prove that $\int_{\tau}\|\mathbf{W}_{n}(t)\|\nu(dt)\stackrel{a.s.}{\rightarrow}0$,
it is enough to show that
$$\int_{\tau}\|\mathbf{W}_{U,n}(t)\|\nu(dt)\stackrel{a.s.}{\rightarrow}0.$$
To this purpose, we observe that the dynamics of $\mathbf{W}_{U,n}=U^\top\mathbf{W}_{n}$
can be derived from~\eqref{eq:urn_dynamics_proofs_2}, so obtaining
$$\mathbf{W}_{U,n}\ =\ [I-r_n(I-\Lambda)]\mathbf{W}_{U,n-1}\ +\ r_nU^\top\Delta \mathbf{M}_{Z,n}\ +\ r_nU^\top\mathbf{R}_{Z,n},$$
where $I$ here indicates an identity matrix of dimension $(d-1)$.
Hence, using Assumption~\ref{assum:2} and
$\E[\Delta\mathbf{M}_{Z,n}\,|\,{\mathcal F}_{n-1}]=0$, we have
\begin{equation*}
\begin{split}
&\E\left[\|\mathbf{W}_{U,n}\|^2|\mathcal{F}_{n-1}\right]
\ = \ \E\left[\overline{\mathbf{W}}_{U,n}^\top\, \mathbf{W}_{U,n}\,|\, {\mathcal F}_{n-1}\right]\ =\\
&\overline{\mathbf{W}}_{U,n-1}^\top\, \mathbf{W}_{U,n-1}-r_n\overline{\mathbf{W}}_{U,n-1}^\top\left(2I-\overline{\Lambda}-\Lambda\right)\mathbf{W}_{U,n-1}
+r_nn^{-\alpha}\psi_{n},
\end{split}
\end{equation*}
where $\{\psi_{n};n\geq1\}$ is a suitable bounded sequence of ${\mathcal F}_{n-1}$-measurable random variables.
Now, since $\mathcal{R}e(\lambda_j(t))<1$ for any $\lambda_j(t)\in Sp(H(t))\setminus\{1\}$ and $t\in\tau$,
the matrix $2I-(\overline{\Lambda}(t)+\Lambda(t))$ is positive definite and hence we can write
\begin{equation*}
\E\left[ \int_{\tau}\|{\mathbf{W}}_{U,n}(t)\|^2\nu(dt)\,|\, {\mathcal F}_{n-1} \right]
\leq \int_{\tau}\|{\mathbf{W}}_{U,n-1}(t)\|^2\nu(dt)\ +\  O(n^{-(1+\alpha)}).
\end{equation*}
Since $\sum_n n^{-(1+\alpha)}<+\infty$, we can conclude
that the real stochastic process $\int_{\tau}\|{\mathbf{W}}_{U,n}(t)\|^2\nu(dt)$ is a positive almost supermartingale and so
it converges almost surely, and in mean since it is also bounded (see \cite{rob-sie}).

In order to prove that the limit is zero, we show the sufficient condition that
$$\E[\int_{\tau}\|\mathbf{W}_{U,n}(t)\|^2\nu(dt)]$$
converges to zero.
To this end, we observe that, from the above computations, we obtain
\begin{equation*}
E[\|\mathbf{W}_{U,n}\|^2]\ \leq\ E[\overline{\mathbf{W}}_{U,n-1}^\top(I-r_n(I-\overline{\Lambda}))(I-r_n(I-\Lambda)) \mathbf{W}_{U,n-1}]\
+\ n^{-(1+\alpha)}C_1
\end{equation*}
for a suitable constant $C_1\geq 0$. Then, we note that the elements
of the diagonal matrix above can be written as follows:
$$[(I-r_n(I-\overline{\Lambda}))(I-r_n(I-\Lambda))]^{jj}\ =\ 1-2r_n(1-{\mathcal R}e(\lambda_j))+r^2_n\, |1-\lambda_j|^2.$$
Setting $a_j(t):=1-{\mathcal R}e(\lambda_j(t))$ and $a^*(t):=\min_{j>1} a_j(t)$,
we have that
\[\begin{aligned}
&\E[\overline{\mathbf{W}}_{U,n-1}^\top
(I-r_n(I-\overline{\Lambda}))(I-r_n(I-\Lambda))\mathbf{W}_{U,n-1}]\ \leq\\
&\sum_{j=2}^{N}(1-2a_j r_n)\E[\overline{W}^j_{U,n-1}W^j_{U,n-1}]\ +C_2 n^{-(1+\alpha)}\ \leq\\
&(1-2a^* r_n)\E[\|\mathbf{W}_{U,n}\|^2]\ +C_2 n^{-(1+\alpha)},
\end{aligned}\]
for a suitable constant $C_2\geq 0$.
Since for any $t\in\tau$ $\max_{j>1}{\mathcal R}e(\lambda_j(t))<1$,
for any $\epsilon>0$ there exists $\delta>0$ such that $\nu(A_{\delta})>1-\epsilon$,
where $A_{\delta}:=\{t\in\tau,a^*(t)>\delta\}$.
Denoting $q_{\delta,n}:=\E[\int_{A_{\delta}}\|\mathbf{W}_{U,n}(t)\|^2\nu(dt)]$,
we have
\begin{equation}\label{eq:final_as_limit}
q_{\delta,n}\ \leq\ (1-2\delta r_n)q_{\delta,n-1}\ +\ (C_1+C_2)n^{-(1+\alpha)},
\end{equation}
which implies $\lim_nq_{\delta,n}=0$ (see \cite{cri-dai-lou-min}).
Hence, for any $\epsilon>0$ we have proved
$$\E\left[\int_{\tau}\|\mathbf{W}_{U,n}(t)\|^2\nu(dt)\right]\ \leq\ \epsilon\ +\ q_{\delta,n}\ \rightarrow\ \epsilon. $$
This concludes the proof of part (a).\\

Concerning part (b), consider the decomposition
$(\mathbf{N}_{t,n}/\Tr(\mathbf{N}_{t,n})-\mathbf{v}(t))=(\mathbf{A}_{1,n}(t)+\mathbf{A}_{2,n}(t))$,
where
\begin{equation*}\begin{aligned}
\mathbf{A}_{1,n}(t) &&:=&\ \frac{\sum_{i=1}^n\ind_{\{T_i=t\}}(\bar{\mathbf{X}}_i-\mathbf{Z}_{i-1}(T_i))}{\sum_{j=1}^n\ind_{\{T_j=t\}}},\\
\mathbf{A}_{2,n}(t) &&:=&\ \frac{\sum_{i=1}^n\ind_{\{T_i=t\}}(\mathbf{Z}_{i-1}(t)-\mathbf{v}(t))}{\sum_{j=1}^n\ind_{\{T_j=t\}}}.
\end{aligned}\end{equation*}
First, using {\cite[Theorem 1]{Chen78}} and the assumption $\sum_{j=1}^n\mu_{j-1}(\{t\})\stackrel{a.s.}{\rightarrow}\infty$,
it follows that $\sum_{j=1}^n\ind_{\{T_j=t\}}\stackrel{a.s.}{\rightarrow}\infty$.
Hence, we can write that, for any $n_0\geq1$,
$$\limsup_{n\rightarrow\infty}\|\mathbf{A}_{2,n}(t)\|\leq \sup_{i\geq n_0}\|\mathbf{Z}_{i-1}(t)-\mathbf{v}(t)\|.$$
Then, $\|\mathbf{A}_{2,n}(t)\|\stackrel{a.s.}{\rightarrow}0$ as a consequence of part (a).
To deal with the term $\mathbf{A}_{1,n}(t)$, consider the martingale process $\{\tilde{\mathbf{A}}_{1,n}(t);n\geq1\}$ defined as follows:
$$\tilde{\mathbf{A}}_{1,n}(t)\ :=\ \sum_{i=1}^n\frac{\ind_{\{T_i=t\}}(\bar{\mathbf{X}}_i-\mathbf{Z}_{i-1}(T_i))}{\sum_{j=1}^i\ind_{\{T_j=t\}}},$$
and notice that $\tilde{\mathbf{A}}_{1,n}(t)$ converges a.s. since with probability one
its bracket process is bounded: $\sum_{i=1}^{\infty}\E[\|\Delta\tilde{\mathbf{A}}_{1,i}(t)\|^2|\F_{i-1}]\leq d\sum_{i=1}^{\infty} i^{-2}<\infty$.
Then, applying the Ces\`{a}ro Lemma it follows that $\|\mathbf{A}_{1,n}(t)\|\stackrel{a.s.}{\rightarrow}0$.

Concerning part (c), consider the decomposition
$(\mathbf{N}_{n}/n-\int_{\tau}\mu(dt)\mathbf{v}(t))=(\mathbf{B}_{1,n}+\mathbf{B}_{2,n}+\mathbf{B}_{3,n})$,
where
\begin{equation*}\begin{aligned}
\mathbf{B}_{1,n} &&:=&\ n^{-1}\sum_{i=1}^n(\bar{\mathbf{X}}_i-\int_{\tau}\mu_{i-1}(dt)\mathbf{Z}_{i-1}(t)),\\
\mathbf{B}_{2,n} &&:=&\ n^{-1}\sum_{i=1}^n\int_{\tau}(\mu_{i-1}(dt)-\mu(dt))\mathbf{Z}_{i-1}(t),\\
\mathbf{B}_{3,n} &&:=&\ n^{-1}\sum_{i=1}^n\int_{\tau}(\mathbf{Z}_{i-1}(t)-\mathbf{v}(t))\mu(dt).
\end{aligned}\end{equation*}
To deal with the term $\mathbf{B}_{1,n}$, consider the martingale process $\{\tilde{\mathbf{B}}_{1,n}(t);n\geq1\}$ defined as follows:
$$\tilde{\mathbf{B}}_{1,n}\ :=\ \sum_{i=1}^ni^{-1}(\bar{\mathbf{X}}_i-\int_{\tau}\mu_{i-1}(dt)\mathbf{Z}_{i-1}(t)),$$
and notice that $\tilde{\mathbf{B}}_{1,n}$ converges a.s. since with probability one
its bracket process is bounded: $\sum_{i=1}^{\infty}\E[\|\Delta\tilde{\mathbf{B}}_{1,i}\|^2|\F_{i-1}]\leq d\sum_{i=1}^{\infty} i^{-2}<\infty$.
Then, applying the Ces\`{a}ro Lemma it follows that $\|\mathbf{B}_{1,n}\|\stackrel{a.s.}{\rightarrow}0$.
Notice that, for any $n_0\geq1$,
$$\limsup_{n\rightarrow\infty}\|\mathbf{B}_{2,n}\|\leq \sup_{i\geq n_0}\int_{\tau}|\mu_{i-1}(dt)-\mu(dt)|,$$
and hence $\|\mathbf{B}_{2,n}\|\stackrel{a.s.}{\rightarrow}0$ using assumption $\int_{\tau}\|\mu_{i-1}(dt)-\mu(dt)\|\stackrel{a.s.}{\rightarrow}0$.
Finally, from part (a) the third term $\|\mathbf{B}_{3,n}\|$ converges to zero a.s. by the Bounded Convergence Theorem.
This concludes the proof.
\endproof

\subsection{Proof of the second-order asymptotic results}\label{subsection_proofs_second_order_properties}

This section contains the proofs of the central limit theorems (CLTs) presented in Section~\ref{section_assumptions_results},
namely Theorem~\ref{thm:second_order_result_strata} and Theorem~\ref{thm:second_order_result_adjusted}.
The key idea of these proofs consists in revisiting the functional urn dynamics in the stochastic approximation (SA) framework,
in the same spirit of the recent works~\cite{AleGhi16,LarPag,zha16}.
For this reason, we now show some basic tools of SA.
The general theory can be found in~\cite{BMP,Duf2,KusYin} (cf. {\cite[Theorem A.2]{LarPag} and \cite[Appendix A]{zha16}})
with different group of conditions.

Consider an $\mathcal{F}_{n}$-measurable multivariate process $\{\mathbf{W}_{n};n\geq1\}$ which evolves as follows:
\begin{equation}\label{SAP}
\forall\, n\geq1,\quad \Delta\mathbf{W}_{n}=-\frac{1}{n}f(\mathbf{W}_{n-1})+\frac{1}{n}(\Delta \mathbf{M}_{n}+\mathbf{R} _{n}),
\end{equation}
where $f$ is a differentiable function,
$\Delta \mathbf{M}_{n}$ is an $\mathcal{F}_{n-1}$-martingale increment and $\mathbf{R}_{n}$ is a remainder term.
Then, assuming that
$$\mathbf{R}_n\stackrel{a.s.}{\longrightarrow} \mathbf{0} \quad \mbox{and} \quad
\sup_{n\geq 1}\mathbb{E}\left[\|\Delta \mathbf{M}_{n}\|^2\,|\,\mathcal{F}_{n-1}\right]<\infty \quad \mbox{a.s.},$$
we have that the set $\mathcal{W}$ of the limiting values of $\mathbf{W}_n$ as $n\rightarrow\infty$ is $a.s.$ a compact connected set, stable by the flow of
$ODE_f\equiv\dot{\mathbf{W}}=-f(\mathbf{W})$.

Moreover, suppose that there exist a constant $\delta>0$ and a deterministic symmetric positive semidefinite matrix $\Gamma$ such that
\begin{equation}\label{HypDM}
	\sup_{n\geq 1}\mathbb{E}[\|\Delta \mathbf{M}_{n}\|^{2+\delta}|\mathcal{F}_{n-1}]<\infty \quad \mbox{a.s.},\qquad
\mathbb{E}\left[\Delta \mathbf{M}_{n}\Delta \mathbf{M}_{n}^\top|\mathcal{F}_{n-1}\right]\stackrel{a.s.}{\longrightarrow}\Gamma,
\end{equation}
and, for any $\epsilon>0$, $n\E[\|\mathbf{R}_n\|^2\ind_{\{\|\mathbf{W}_{n}-\mathbf{W}\|\leq\epsilon\}}]\longrightarrow0$.
Then, considering an equilibrium point $\mathbf{W}$ of $\{\mathbf{w}:f(\mathbf{w})=0\}$ such that all the
eigenvalues of $\mathcal{D}f(\mathbf{W})$ have real parts bigger than 1/2, we have that
$\sqrt{n}(\mathbf{W}_{n}-\mathbf{W})\stackrel{\mathcal{L}}{\longrightarrow}\mathcal{N}(0,\Sigma)$,
where $\Sigma=\int_0^{\infty}e^{u(I/2-\mathcal{D}f(\mathbf{W}))}\Gamma e^{u(I/2-\mathcal{D}f(\mathbf{W}))^\top}du$.

\proof[Proof of Theorem~\ref{thm:second_order_result_strata}]
Initially, we need to express in the SA form~\eqref{SAP} the joint dynamics of the following processes:
\begin{itemize}
\item[(1)] the urn proportion in correspondence of all the covariate profiles,
$$\mathbf{Z}_n:=(\mathbf{Z}_n(t),t\in\tau)^\top;$$
\item[(2)] the proportion of subjects of all covariate profiles assigned to the treatments,
$$\tilde{\mathbf{N}}_n:=(\tilde{\mathbf{N}}_{t,n},t\in\tau)^\top,\qquad
\mbox{where}\qquad \tilde{\mathbf{N}}_{t,n}:=\frac{\mathbf{N}_{t,n}}{\Tr(\mathbf{N}_{t,n})};$$
\item[(3)] the adaptive estimators of features of interest related with the response distributions conditioned on
each covariate profile, $$\hat{\mathbf{\theta}}_{n}:=(\hat{\mathbf{\theta}}_{t,n},t\in\tau)^\top,\qquad
\mbox{where}\qquad \hat{\mathbf{\theta}}_{t,n}:=(\hat{\mathbf{\theta}}^j_{t,n},j\in\{1,..,d\})^\top;$$
\item[(4)] the proportion of subjects with all covariate profiles observed in the trial,
$$\mathbf{Q}_n:=({Q}_{t,n},t\in\tau)^\top,\qquad
\mbox{where}\qquad Q_{t,n}:=\frac{\Tr(\mathbf{N}_{t,n})}{n}=\frac{1}{n}\sum_{i=1}^n\ind_{\{T_i=t\}}.$$
\end{itemize}
Then, the CLT follows by applying to this joint dynamics the standard theory of the SA.

First using~\eqref{eq:urn_dynamics}, we express the joint dynamics of $(\mathbf{Z}_n(t),\mathbf{N}_{t,n})$ as follows:
let $\mathbf{N}_0=\mathbf{0}$ and $\mathbf{Y}_0=\mathbf{1}$, and for any $n\geq 0$
\begin{equation}\label{eq:dynamics_proofs_1_strata}
\left\{\begin{aligned}
&\mathbf{Y}_n(t)\ =\ \mathbf{Y}_{n-1}(t)\ +\ D_n(t)\mathbf{X}_n(t),\\
&\mathbf{N}_{t,n}\ =\ \mathbf{N}_{t,n-1}\ +\ \bar{\mathbf{X}}_n\ind_{\{T_n=t\}}.
\end{aligned}\right.\end{equation}
Notice that, by defining $r_n:=(d+n)^{-1}$ and using Assumption~\ref{assum:1},
$r_n\mathbf{Y}_{n}=\mathbf{Y}_{n}/\Tr(\mathbf{Y}_{n})=\mathbf{Z}_{n}$.
Then in~\eqref{eq:dynamics_proofs_1_strata}, if we multiply the dynamics of $\mathbf{Y}_n(t)$ by $r_n$ and
the dynamics of $\mathbf{N}_{t,n}$ by $\Tr(\mathbf{N}_{t,n})^{-1}$, we obtain
\begin{equation}\label{eq:dynamics_proofs_2_strata}
\left\{\begin{aligned}
&\mathbf{Z}_n(t)-\mathbf{Z}_{n-1}(t)\ =\ -r_n(\mathbf{Z}_{n-1}(t)-D_n(t)\mathbf{X}_n(t)),\\
&\frac{\mathbf{N}_{t,n}}{\Tr(\mathbf{N}_{t,n})}-\frac{\mathbf{N}_{t,n-1}}{\Tr(\mathbf{N}_{t,n-1})}\ =\
-\frac{\ind_{\{T_n=t\}}}{\Tr(\mathbf{N}_{t,n})}\left(\frac{\mathbf{N}_{t,n-1}}{\Tr(\mathbf{N}_{t,n-1})}-\bar{\mathbf{X}}_n\right),
\end{aligned}\right.\end{equation}
where in~\eqref{eq:dynamics_proofs_2_strata} we have used the relations $r_nr^{-1}_{n-1}=(1-r_n)$ and
$$\frac{\Tr(\mathbf{N}_{t,n-1})}{\Tr(\mathbf{N}_{t,n})}=(1-\ind_{\{T_n=t\}}\Tr(\mathbf{N}_{t,n})^{-1}).$$
Then, recalling $\tilde{\mathbf{N}}_{t,n}=\mathbf{N}_{t,n}/\Tr(\mathbf{N}_{t,n})$ and adding to~\eqref{eq:dynamics_proofs_2_strata}
the dynamics of $\{\hat{\mathbf{\theta}}^j_{t,n};n\geq n_0\}$ expressed in~\eqref{SAP_theta_strata},
we obtain
\begin{equation}\label{eq:dynamics_proofs_3_strata}
\left\{\begin{aligned}
&\Delta\mathbf{Z}_n(t)\ =\ -r_n(\mathbf{Z}_{n-1}(t)-D_n(t)\mathbf{X}_n(t)),\\
&\Delta\tilde{\mathbf{N}}_{t,n}\ =\
-\frac{\ind_{\{T_n=t\}}}{\Tr(\mathbf{N}_{t,n})}\left(\tilde{\mathbf{N}}_{t,n-1}-\bar{\mathbf{X}}_n\right),\\
&\Delta\hat{\mathbf{\theta}}^j_{t,n}\ =\ -\frac{\bar{X}_n^j\ind_{\{T_n=t\}}}{N^j_{t,n}}
(f_{t,j}(\hat{\mathbf{\theta}}^j_{t,n-1})-\Delta \mathbf{M}_{t,j,n}-\mathbf{R}_{t,j,n}).
\end{aligned}\right.\end{equation}
Now, let $Q_{t,n}:=\Tr(\mathbf{N}_{t,n})/n$, where by assumption $f_{\mu,t}(\cdot)\geq\epsilon>0$ we have
$\liminf_nQ_{t,n}\geq$ $\liminf_n \mu_n(t)\geq\epsilon>0$ with probability one.
Notice that
$$r_n^{-1}\frac{\ind_{\{T_n=t\}}}{\Tr(\mathbf{N}_{t,n})}\ =\ r_n^{-1}\frac{\ind_{\{T_n=t\}}}{\Tr(\mathbf{N}_{t,n-1})+1}\ =\
\frac{\ind_{\{T_n=t\}}}{Q_{t,n-1}}+\frac{\psi_{\theta^j_t,n}}{n},$$
and analogously,
$$r_n^{-1}\frac{\bar{X}_n^j\ind_{\{T_n=t\}}}{N^j_{t,n}}\ =\ r_n^{-1}\frac{\bar{X}_n^j\ind_{\{T_n=t\}}}{N^j_{t,n-1}+1}\ =\
\frac{\bar{X}_n^j\ind_{\{T_n=t\}}}{\tilde{N}^j_{t,n-1}Q_{t,n-1}}+\frac{\psi_{N_t,n}}{n},$$
where $\{\psi_{N_t,n};n\geq1\}$ and $\{\psi_{\theta^j_t,n};n\geq1\}$ are suitable bounded sequence of ${\mathcal F}_{n}$-measurable random variables.
Then, using the above relations in~\eqref{eq:dynamics_proofs_3_strata} we obtain
\begin{equation}\label{eq:dynamics_proofs_4_strata}
\left\{\begin{aligned}
&\Delta\mathbf{Z}_n(t)\ =\ -r_n(\mathbf{Z}_{n-1}(t)-D_n(t)\mathbf{X}_n(t)),\\
&\Delta\tilde{\mathbf{N}}_{t,n}\ =\
-r_n\frac{\ind_{\{T_n=t\}}}{Q_{t,n-1}}\left(\tilde{\mathbf{N}}_{t,n-1}-\bar{\mathbf{X}}_n\right)+r_n\mathbf{R}_{N_t,n},\\
&\Delta\hat{\mathbf{\theta}}^j_{t,n}\ =\ -r_n\frac{\bar{X}_n^j\ind_{\{T_n=t\}}}{\tilde{N}^j_{t,n-1}Q_{t,n-1}}
(f_{t,j}(\hat{\mathbf{\theta}}^j_{t,n-1})-\Delta \mathbf{M}_{t,j,n})+r_n\mathbf{R}_{\theta_t^j,n},
\end{aligned}\right.\end{equation}
where $\mathbf{R}_{N_t,n},\mathbf{R}_{\theta_t^j,n}\in{\mathcal F}_{n}$ are
suitable random variables that converges to zero a.s. and
$(\E[\|\mathbf{R}_{N_t,n}\|^2]+\E[\|\mathbf{R}_{\theta_t^j,n}\|^2])=o(n)$.
Now, in order to express the dynamics in~\eqref{eq:dynamics_proofs_4_strata} in the SA form~\eqref{SAP},
we need also to consider the process $\{{Q}_{t,n};n\geq1\}$ and
to rewrite~\eqref{eq:dynamics_proofs_4_strata} as follows:
\begin{equation*}
\left\{\begin{aligned}
&\Delta\mathbf{Z}_n(t)\ =\ -r_nf_{Z,t}(\mathbf{Z}_{n-1}(t))\ +\ r_n\Delta \mathbf{M}_{Z(t),n},\\
&\Delta\tilde{\mathbf{N}}_{t,n}\ =\
-r_nf_{N,t}(\mathbf{Z}_{n-1}(t),\tilde{\mathbf{N}}_{t,n-1},\hat{\theta}_{t,n-1},Q_{t,n-1})\ +\
r_n(\Delta \mathbf{M}_{N_t,n}+r_n\mathbf{R}_{N_t,n}),\\
&\Delta\hat{\mathbf{\theta}}^j_{t,n}\ =\ -r_n
f_{\theta,t}(\mathbf{Z}_{n-1}(t),\tilde{\mathbf{N}}_{t,n-1},\hat{\theta}_{t,n-1},Q_{t,n-1})
+r_n(\Delta \mathbf{M}_{\theta_t^j,n}+\mathbf{R}_{\theta_t^j,n}),\\
&\Delta{Q}_{t,n}\ =\
-r_nf_{Q,t}(\tilde{\mathbf{N}}_{t,n-1},\hat{\theta}_{t,n-1},Q_{t,n-1})\ +\ r_n\Delta {M}_{Q_t,n},
\end{aligned}\right.\end{equation*}
where
\[\begin{aligned}
&f_{Z,t}(\mathbf{Z}_{n-1}(t)):=(I-H(t))\mathbf{Z}_{n-1}(t)+\mathbf{v}(t)(\mathbf{1}^\top\mathbf{Z}_{n-1}(t)-1),\\
&f_{N,t}(\mathbf{Z}_{n-1}(t),\tilde{\mathbf{N}}_{t,n-1},\hat{\theta}_{t,n-1},Q_{t,n-1}):=
    \frac{\mu_{n-1}(t)}{Q_{t,n-1}}\left(\tilde{\mathbf{N}}_{t,n-1}-\mathbf{Z}_{n-1}(t)\right),\\
&f_{\theta,t}(\mathbf{Z}_{n-1}(t),\tilde{\mathbf{N}}_{t,n-1},\hat{\theta}_{t,n-1},Q_{t,n-1}):=
\frac{\mu_{n-1}(t)Z^j_{n-1}(t)}{\tilde{N}^j_{t,n-1}Q_{t,n-1}}
f_{t,j}(\hat{\mathbf{\theta}}^j_{t,n-1}),\\
&f_{M,t}(\tilde{\mathbf{N}}_{t,n-1},\hat{\theta}_{t,n-1},Q_{t,n-1}):=
(Q_{t,n-1}-\mu_{n-1}(t)),
\end{aligned}\]
and
\[\begin{aligned}
\Delta \mathbf{M}_{Z(t),n}&&:=&(D_n\mathbf{X}_n-H\mathbf{Z}_{n-1})(t),\\
\Delta \mathbf{M}_{N_t,n}\ &&:=&\
(\ind_{\{T_n=t\}}-\mu_{n-1}(t))
\frac{(\tilde{\mathbf{N}}_{t,n-1}-\mathbf{Z}_{n-1}(t))}{Q_{t,n-1}}-\ \frac{\ind_{\{T_n=t\}}}{Q_{t,n-1}}(\bar{\mathbf{X}}_{n}-\mathbf{Z}_{n-1}(t)),\\
\Delta \mathbf{M}_{\theta_t^j,n}\ &&:=&\ (\bar{X}_n^j\ind_{\{T_n=t\}}-\mu_{n-1}(t)Z^j_{n-1}(t))
    \frac{f_{t,j}(\hat{\mathbf{\theta}}^j_{t,n-1})}{\tilde{N}^j_{t,n-1}Q_{t,n-1}}
    + \frac{\bar{X}_n^j\ind_{\{T_n=t\}}}{\tilde{N}^j_{t,n-1}Q_{t,n-1}}\Delta \mathbf{M}_{t,j,n},\\
\Delta M_{Q,n}(t)&&:=&(\ind_{\{T_n=t\}}-\mu_{n-1}(t)),
\end{aligned}\]
are martingale increments since $\E[D_n(t)|T_n,\mathbf{\bar{X}}_n]=H(t)$,
$\E[\bar{\mathbf{X}}_n|\mathcal{F}_{n-1},T_n]=\mathbf{Z}_{n-1}(T_n)$,
$\E[\Delta \mathbf{M}_{t,j,n}|\mathcal{F}_{n-1},T_n,\bar{\mathbf{X}}_n]=0$,
$\E[\ind_{\{T_n=t\}}|\mathcal{F}_{n-1}]=\mu_{n-1}(t)$.

Let us now introduce the joint processes $\{\mathbf{W}_n,n\geq1\}$ defined as
$\mathbf{W}_n:=(\mathbf{Z}_n,\tilde{\mathbf{N}}_{n},\hat{\mathbf{\theta}}_n,\mathbf{Q}_{n})^\top$,
and note that its dynamics can be expressed in the SA form~\eqref{SAP} as follows:
\begin{equation}\label{eq:dynamics_proofs_6_strata}
\Delta\mathbf{W}_n\ =\ -r_nf_{W}(\mathbf{W}_{n-1})\ +\ r_n(\Delta\mathbf{M}_{W,n}+\mathbf{R}_{W,n}),
\end{equation}
where
\begin{itemize}
\item[(i)] $f_{W}:=(f_{Z},f_{N},f_{\theta},f_{Q})^\top$, where $f_{Z}:=(f_{Z(t)},t\in\tau)^\top$, $f_{N}:=(f_{N_t},t\in\tau)^\top$,
$f_{\theta}:=(f_{\theta^j_t},t\in\tau,j\in\{1,..,d\})^\top$, $f_{Q}:=(f_{Q_t},t\in\tau)^\top$;
\item[(ii)] $\Delta\mathbf{M}_{W,n}:=(\Delta\mathbf{M}_{Z,n},\Delta\mathbf{M}_{N,n},\Delta\mathbf{M}_{\theta,n},\Delta\mathbf{M}_{Q,n})^\top$,
where\\ $\Delta\mathbf{M}_{Z,n}:=(\Delta\mathbf{M}_{Z(t),n},t\in\tau)^\top$, $\Delta\mathbf{M}_{N,n}:=(\Delta\mathbf{M}_{N_t,n},t\in\tau)^\top$, \\ $\Delta\mathbf{M}_{\theta,n}:=(\Delta\mathbf{M}_{\theta^j_t,n},t\in\tau,j\in\{1,..,d\})^\top$, $\Delta\mathbf{M}_{Q,n}:=(\Delta\mathbf{M}_{Q_t,n},t\in\tau)^\top$;
\item[(iii)] $\mathbf{R}_{W,n}:=(\mathbf{0},\mathbf{R}_{N,n},\mathbf{R}_{\theta,n},\mathbf{0})^\top$, where
$\mathbf{R}_{N,n}:=(\mathbf{R}_{N_t,n},t\in\tau)^\top$,\\ $\mathbf{R}_{\theta,n}:=(\mathbf{R}_{\theta^j_t,n},t\in\tau,j\in\{1,..,d\})^\top$.
\end{itemize}
Since $\mathbf{R}_{W,n}\stackrel{a.s.}{\longrightarrow}0$ and, using~\eqref{ass:martingale_theta_finite}
in~\ref{assum:6a}, $\sup_n\E[\|\Delta\mathbf{M}_{W,n}\|^2]<\infty$,
we have that the set $\mathcal{W}$ of the limiting values of $\mathbf{W}_n$ is a stable set by the flow of $\dot{\mathbf{W}}=-f_W(\mathbf{W})$.
Notice that the set $\{\mathbf{w}:f_W(\mathbf{w})=\mathbf{0}\}$ is composed only of the element
$\mathbf{W}:=(\mathbf{v},\mathbf{v},\mathbf{\theta},\mu)^\top$, where
$\mathbf{v}:=(\mathbf{v}(t),t\in\tau)^\top$ and $\mu:=(\mu(t),t\in\tau)^\top$.
Moreover, we recall from Theorem~\ref{thm:first_order_result} that we have $\mathbf{Z}_n\stackrel{a.s.}{\longrightarrow}\mathbf{v}$ and $\tilde{\mathbf{N}}_{n}\stackrel{a.s.}{\longrightarrow}\mathbf{v}$, and by~\ref{assum:6a},
we have $\hat{\mathbf{\theta}}_n\stackrel{a.s.}{\longrightarrow}\mathbf{\theta}$.
Since $\mathbf{\mu}_{n}(t)=f_{\mu,t}(\tilde{\mathbf{N}}_{t,n},\hat{\mathbf{\theta}}_{t,n})
\stackrel{a.s.}{\longrightarrow}f_{\mu,t}(\mathbf{v}(t),\mathbf{\theta}_t)=\mathbf{\mu}(t)$ and
$\mathbf{Q}_{n}-\sum_{i=1}^n\mathbf{\mu}_{i-1}/n=\sum_{i=1}^n\Delta\mathbf{M}_{Q,n}/n\stackrel{a.s.}{\longrightarrow}\mathbf{0}$,
we also have $\mathbf{Q}_{n}\stackrel{a.s.}{\longrightarrow}\mathbf{\mu}$, which implies $\mathbf{W}_n\stackrel{a.s.}{\longrightarrow}\mathbf{W}$.

In order to show the existence of a stable attracting area which contains a neighborhood of $\mathbf{W}$, 
it is sufficient (see \cite[p. 1077]{FortPages}) to show that $\{\Re e(Sp(\mathcal{D}f_{W}(\mathbf{W})))>0\}$, where
\begin{equation}\label{eq:matrix_Df_proof_strata}
\mathcal{D}f_{W}(\mathbf{W})=
\begin{pmatrix}
\mathcal{D}_Zf_{Z}(\mathbf{W}) & 0 & 0 & 0\\
-I & I & 0 & 0\\
0 & 0 & \mathcal{D}_{\theta}f_{\theta}(\mathbf{W}) & 0\\
0 & \mathcal{D}_Nf_{Q}(\mathbf{W}) & \mathcal{D}_{\theta}f_{Q}(\mathbf{W}) & I
\end{pmatrix},
\end{equation}
and all the terms in~\eqref{eq:matrix_Df_proof_strata} are block-diagonal matrices, whose $t^{th}$ block is: $[\mathcal{D}_Zf_{Z}(\mathbf{W})]^{tt}=(I-H(t)+\mathbf{v}(t)\mathbf{1}^\top)$,
$[\mathcal{D}_{\theta}f_{\theta}(\mathbf{W})]^{tt}=diag(\mathcal{D}f_{t,j}(\mathbf{\theta}_t^j),j\in\{1,..,d\})$,
$[\mathcal{D}_Nf_{Q}(\mathbf{W})]^{tt}=\mathcal{D}_Nf_{\mu,t}(\mathbf{W})$ and
$[\mathcal{D}_Nf_{Q}(\mathbf{W})]^{tt}=\mathcal{D}_{\theta}f_{\mu,t}(\mathbf{W})$.
Note from the structure of $\mathcal{D}f_{W}(\mathbf{W})$ in~\eqref{eq:matrix_Df_proof_strata}
that $\{\Re e(Sp(\mathcal{D}f_{W}(\mathbf{W})))>0\}$ follows by establishing that for any $t\in\tau$ and $j\in\{1,..,d\}$
$$\{\Re e(Sp(I-H(t)+\mathbf{v}(t)\mathbf{1}^\top))>0\}\ \mbox{ and }
\{\Re e(Sp(\mathcal{D}f_{{t,j}}(\mathbf{\theta}_t^j)))>0\}.$$
Since $(I-H(t))=V(t)(I-\Lambda(t))U(t)$ from~\eqref{eq:relations} and~\eqref{eq:decomposition_H_star}, we have that
$$Sp(I-H(t)+\mathbf{v}(t)\mathbf{1}^\top)=\{1\}\cup\{1-\lambda(t),\lambda(t)\in Sp(H(t))\setminus\{1\}\}.$$
Then $\{\Re e(Sp(\mathcal{D}f_{W}(\mathbf{W})))>0\}$ follows by $\{\max_{t\in\tau}{\mathcal Re}(\lambda_H^{*}(t))<1/2\}$ from~\ref{assum:5}
and $\{\min_{t\in\tau}{\mathcal Re}(\lambda^{*}_{\theta_t^j})>1/2\}$ from~\ref{assum:6a}.

We now show that the assumptions of the CLT for processes in the SA form are satisfied by
the dynamics in~\eqref{eq:dynamics_proofs_6_strata} of the joint process $\{\mathbf{W}_{n},n\geq1\}$.
First, note that using the above arguments we obtain $\{\Re e(Sp(\mathcal{D}f_{W}(\mathbf{W})))>1/2\}$.
Then, it is immediate to see that $\E[\|\mathbf{R}_{W,n}\|^2]=o(n)$ and
the first condition in~\eqref{HypDM} is satisfied using~\eqref{ass:martingale_theta_finite} in Assumption~\ref{assum:6a}.
Concerning the second condition in~\eqref{HypDM}, we need to show that there exists a deterministic symmetric positive
semidefinite matrix $\Gamma$ such that
$$\E[\Delta\mathbf{M}_{W,n}(\Delta\mathbf{M}_{W,n})^\top|\F_{n-1}]\stackrel{a.s.}{\longrightarrow}\Gamma=
\begin{pmatrix}
\Gamma_{ZZ} & \Gamma_{ZN} & \Gamma_{Z\theta} & \Gamma_{ZQ}\\
\Gamma_{ZN}^\top & \Gamma_{NN} & \Gamma_{N\theta} & \Gamma_{N Q}\\
\Gamma_{Z\theta}^\top & \Gamma_{N\theta}^\top & \Gamma_{\theta \theta} & \Gamma_{\theta Q}\\
\Gamma_{ZQ}^\top & \Gamma_{NQ}^\top & \Gamma_{\theta Q}^\top & \Gamma_{QQ}
\end{pmatrix}.$$
First, note that since $(\tilde{\mathbf{N}}_{t,n-1}-\mathbf{Z}_{n-1}(t))\stackrel{a.s.}{\longrightarrow}0$
and $f_{t,j}(\hat{\mathbf{\theta}}^j_{t,n-1})\stackrel{a.s.}{\longrightarrow}0$,
these terms do not contribute to $\Gamma$; hence in the following calculations
they will be omitted by $\Delta \mathbf{M}_{N_t,n}$ and $\Delta \mathbf{M}_{\theta_t^j,n}$, respectively.
Moreover, let us introduce for any $t,s\in\tau$ and $j\in\{1,..,d\}$,
a vector $\mathbf{g}(t,s,\mathbf{e}_j)\in\mathcal{S}$ such that $g^k(t,s,\mathbf{e}_j)$, $k\in\{1,..,d\}$, is defined as follows:
\begin{equation}\label{def:vector_g}
\frac{\left(\ \min\left\{\sum_{i=1}^{k}v^{i}(t);\sum_{i=1}^{j}v^{i}(s)\right\}\ -\
\max\left\{\sum_{i=1}^{k-1}v^{i}(t);\sum_{i=1}^{j-1}v^{i}(s)\right\}\ \right)^+}{v^{j}(s)}.
\end{equation}
Then, before computing the terms in $\Gamma$ we show that for any $t\in\tau$
\begin{equation}\label{eq:convergence_X_g}
\E[\|\mathbf{X}_n(t)-\mathbf{g}(t_1,T_n,\mathbf{X}_n)\||\F_{n-1},T_n,\bar{\mathbf{X}}_n]\stackrel{a.s.}{\longrightarrow}0,
\end{equation}
To this end, first note from~\eqref{def:X_n_statistical_part_simply} that $\mathbf{X}_n(t)$ is a continuous function of $\{\mathbf{Z}_{n-1}(s),s\in\tau\}$
conditioned on $\F_{n-1}$, $T_n$ and $\bar{\mathbf{X}}_n$;
then~\eqref{eq:convergence_X_g} follows by Theorem~\ref{thm:first_order_result}
which states $\mathbf{Z}_{n-1}(t)\stackrel{a.s.}{\longrightarrow}\mathbf{v}(t)$ for any $t\in\tau$,
since $\tau$ has a finite number of elements.
We now compute the terms in $\Gamma$.

\textit{Computation of $\Gamma_{ZZ}:=a.s.-\lim_n\E[\Delta\mathbf{M}_{Z,n}(\Delta\mathbf{M}_{Z,n})^\top|\F_{n-1}]$}.
For any $t_1,t_2\in\tau$, we have
\[\begin{aligned}
&\E[\Delta\mathbf{M}_{Z(t_1),n}(\Delta\mathbf{M}_{Z(t_2),n})^\top|\F_{n-1}]
=\E[D_n(t_1)\mathbf{X}_n(t_1)(\Delta\mathbf{M}_{Z(t_2),n})^\top|\F_{n-1}]\\
&=\E[D_n(t_1)\mathbf{X}_n(t_1)\mathbf{X}_n^\top(t_2)D_n^\top(t_2)|\F_{n-1}]-H(t_1)\mathbf{Z}_{n-1}(t_1)\mathbf{Z}_{n-1}^\top(t_2)H^\top(t_2)
\end{aligned}\]
Consider the decomposition $D_n(t_1)\mathbf{X}_n(t_1)\mathbf{X}_n^\top(t_2)D_n^\top(t_2)=(B_{1n}+B_{2n})$,
where
\[\begin{aligned}
B_{1n}&&:=&\ D_n(t_1)(\mathbf{X}_n(t_1)\mathbf{X}_n^\top(t_2)-
\mathbf{g}(t_1,T_n,\bar{\mathbf{X}}_n)\mathbf{g}^\top(t_2,T_n,\bar{\mathbf{X}}_n))D_n^\top(t_2)\\
B_{2n}&&:=&\ D_n(t_1)\mathbf{g}(t_1,T_n,\bar{\mathbf{X}}_n)\mathbf{g}^\top(t_2,T_n,\bar{\mathbf{X}}_n)D_n^\top(t_2).
\end{aligned}\]
Using~\eqref{eq:convergence_X_g} and since $D_n$ is a.s. bounded, it follows by the the Dominated Convergence Theorem that $\E[B_{1n}|\F_{n-1}]\stackrel{a.s.}{\longrightarrow}0$.
In addition, since the probability distribution of the random variables in $B_{2n}$, i.e. $(T_n,\bar{\mathbf{X}}_n,\bar{\xi}_n)$, conditioned on $\F_{n-1}$,
converges a.s. as $n$ increases to infinity,
we obtain $\E[B_{2n}|\F_{n-1}]\stackrel{a.s.}{\longrightarrow}\E[D(t_1)\mathbf{g}(t_1,T,\bar{\mathbf{X}})\mathbf{g}^\top(t_2,T,\bar{\mathbf{X}})D^\top(t_2)]$.
Hence, we have proved the following:
$$\Gamma_{ZZ}^{t_1t_2}:=\E[D(t_1)\mathbf{g}(t_1,T,\bar{\mathbf{X}})\mathbf{g}^\top(t_2,T,\bar{\mathbf{X}})D^\top(t_2)]-\mathbf{v}(t_1)\mathbf{v}^\top(t_2).$$

\textit{Computation of $\Gamma_{NN}:=a.s.-\lim_n\E[\Delta\mathbf{M}_{N,n}(\Delta\mathbf{M}_{N,n})^\top|\F_{n-1}]$}.
Note that $\E[\Delta\mathbf{M}_{N_{t_1},n}(\Delta\mathbf{M}_{N_{t_2},n})^\top|\F_{n-1}]=0=\Gamma_{NN}^{t_1t_2}$ for any $t_1\neq t_2$,
while for $t_1=t_2=t$ we have
\[\begin{aligned}
&\E[\Delta\mathbf{M}_{N_{t},n}(\Delta\mathbf{M}_{N_{t},n})^\top|\F_{n-1}]\\
&=Q_{t,n-1}^{-2}\mu_{n-1}(t)\E[(\bar{\mathbf{X}}_{n}-\mathbf{Z}_{n-1}(t))(\bar{\mathbf{X}}_{n}-\mathbf{Z}_{n-1}(t))^\top|\F_{n-1},T_n=t]\\
&=Q_{t,n-1}^{-2}\mu_{n-1}(t)(diag(\mathbf{Z}_{n-1}(t))-\mathbf{Z}_{n-1}(t)\mathbf{Z}_{n-1}^\top(t))\\
&\stackrel{a.s.}{\longrightarrow}\Gamma_{NN}^{tt}:=\mu^{-1}(t)(diag(\mathbf{v}(t))-\mathbf{v}(t)\mathbf{v}^\top(t)).
\end{aligned}\]

\textit{Computation of $\Gamma_{ZN}:=a.s.-\lim_n\E[\Delta\mathbf{M}_{Z,n}(\Delta\mathbf{M}_{N,n})^\top|\F_{n-1}]$}.
For any $t_1,t_2\in\tau$ we have that
\[\begin{aligned}
&\E[\Delta\mathbf{M}_{Z(t_1),n}(\Delta\mathbf{M}_{N_{t_2},n})^\top|\F_{n-1}]
=\E[D_n(t_1)\mathbf{X}_n(t_1)(\Delta\mathbf{M}_{N_{t_2},n})^\top(t_2)|\F_{n-1}]\\
&=Q_{t_2,n-1}^{-1}\mu_{n-1}(t_2)\left(
\E[D_n(t_1)\mathbf{X}_n(t_1)\bar{\mathbf{X}}_n^\top|\F_{n-1},T_n=t_2]-H(t_1)\mathbf{Z}_{n-1}(t_1)\mathbf{Z}_{n-1}^\top(t_2)\right).
\end{aligned}\]
Note that the above term $\E[D_n(t_1)\mathbf{X}_n(t_1)\bar{\mathbf{X}}_n^\top|\F_{n-1},T_n=t_2]$
can be expressed as follows:
\[\begin{aligned}
&\E[\E[D_n(t_1)|\F_{n-1},T_n=t_2,\bar{\mathbf{X}}_n]\mathbf{X}_n(t_1)\bar{\mathbf{X}}_n^\top|\F_{n-1},T_n=t_2]=\\
&H(t_1)\E[\mathbf{X}_n(t_1)\bar{\mathbf{X}}_n^\top|\F_{n-1},T_n=t_2].
\end{aligned}\]
Then, since using~\eqref{eq:convergence_X_g} it follows by the the Dominated Convergence Theorem that
$$\E[\mathbf{X}_n(t_1)-\mathbf{g}(t_1,t_2,\bar{\mathbf{X}}_n)|\F_{n-1},T_n=t_2]\stackrel{a.s.}{\longrightarrow}0,$$
we can directly consider
$H(t_1)\E[\mathbf{g}(t_1,t_2,\bar{\mathbf{X}}_n)\bar{\mathbf{X}}_n^\top|\F_{n-1},T_n=t_2]$;
then, since the probability distribution of $\bar{\mathbf{X}}_n$ conditioned on $\F_{n-1}$ and $T_n$
converges a.s. as $n$ increases to infinity,
we obtain $$\E[\mathbf{g}(t_1,t_2,\bar{\mathbf{X}}_n)\bar{\mathbf{X}}_n^\top|\F_{n-1},T_n=t_2]
\stackrel{a.s.}{\longrightarrow}\E[\mathbf{g}(t_1,t_2,\bar{\mathbf{X}}_n)\bar{\mathbf{X}}^\top|T=t_2].$$
Hence, we have proved the following:
$$\Gamma_{ZN}^{t_1 t_2}:=H(t_1)G(t_1,t_2)diag(\mathbf{v}(t_2))-\mathbf{v}(t_1)\mathbf{v}^\top(t_2).$$

\textit{Computation of $\Gamma_{\theta\theta}:=a.s.-\lim_n\E[\Delta\mathbf{M}_{\theta,n}(\Delta\mathbf{M}_{\theta,n})^\top|\F_{n-1}]$}.
Since for any $j_1\neq j_2$ or $t_1\neq t_2$ we have $\E[\Delta\mathbf{M}_{\theta_{t_1}^{j_1},n}(\Delta\mathbf{M}_{\theta_{t_2}^{j_2},n})^\top|\F_{n-1}]=0$,
we have that $\Gamma_{\theta\theta}$ is a block-diagonal matrix. In particular, for any $t\in\tau$ we have that
$\Gamma_{\theta\theta}^{tt}=diag([\Gamma_{\theta\theta}^{tt}]^{jj},j\in\{1,..,d\})^\top$, where
\[\begin{aligned}
&\E[\Delta\mathbf{M}_{\theta_t^j,n}(\Delta\mathbf{M}_{\theta_t^j,n})^\top|\F_{n-1}]\\
&=(\tilde{N}^j_{t,n-1}Q_{t,n-1})^{-2}\mu_{n-1}(t)Z^j_{n-1}(t)
\times\E[\Delta\mathbf{M}_{t,j,n}(\Delta\mathbf{M}_{t,j,n})^\top|\F_{n-1},T_n=t,\bar{X}_n^j=1]\\
&\stackrel{a.s.}{\longrightarrow}[\Gamma_{\theta\theta}^{tt}]^{jj}:=(v^j(t)\mu(t))^{-1}\E[\Delta\mathbf{M}_{t,j}(\Delta\mathbf{M}_{t,j})^\top|T=t,\bar{X}^j=1].
\end{aligned}\]

\textit{Computation of $\Gamma_{Z\theta}$}.
For any $t_1,t_2\in\tau$ and $j\in\{1,..,d\}$ we have that
\[\begin{aligned}
&\E[\Delta\mathbf{M}_{Z(t_1),n}(\Delta\mathbf{M}_{\theta_{t_2}^j,n})^\top|\F_{n-1}]
=\E[D_n(t_1)\mathbf{X}_n(t_1)(\Delta\mathbf{M}_{\theta_{t_2}^j,n})^\top|\F_{n-1}]\\
&=(\tilde{N}^j_{t_2,n-1}Q_{t_2,n-1})^{-1}\mu_{n-1}(t_2)Z^j_{n-1}(t_2)
\E[D_n(t_1)\mathbf{X}_n(t_1)(\Delta\mathbf{M}_{{t_2},j,n})^\top|\F_{n-1},T_n=t_2,\bar{X}_n^j=1].
\end{aligned}\]
Then, since using~\eqref{eq:convergence_X_g} it follows by the the Dominated Convergence Theorem that
$\E[\mathbf{X}_n(t_1)-\mathbf{g}(t_1,t_2,\mathbf{e}_j)|\F_{n-1},T_n=t_2,\bar{X}_n^j=1]\stackrel{a.s.}{\longrightarrow}0$,
we can directly consider $$\E[D_n(t_1)\mathbf{g}(t_1,t_2,\mathbf{e}_j)(\Delta\mathbf{M}_{{t_2},j,n})^\top|\F_{n-1},T_n=t_2,\bar{X}_n^j=1];$$
then, since the probability distribution of $\bar{\xi}_n$ conditioned on $\F_{n-1}$, $T_n$ and $\bar{\mathbf{X}}_n$
does not change, we have proved that
$$[\Gamma_{Z\theta}^{t_1t_2}]^{jj}:= \E[D(t_1)\mathbf{g}(t_1,t_2,\mathbf{e}_j)(\Delta\mathbf{M}_{{t_2},j})^\top|T=t_2,\bar{X}^j=1].$$

\textit{Computation of $\Gamma_{N\theta}:=a.s.-\lim_n\E[\Delta\mathbf{M}_{N,n}(\Delta\mathbf{M}_{\theta,n})^\top|\F_{n-1}]$}.
For any $t_1,t_2\in\tau$ we have that
$$\E[\Delta\mathbf{M}_{N_{t_1},n}(\Delta\mathbf{M}_{\theta_{t_2}^j,n})^\top|\F_{n-1}]=
\E[\Delta\mathbf{M}_{N_{t_1},n}\E[(\Delta\mathbf{M}_{\theta_{t_2}^j,n})^\top |\F_{n-1},T_n,\bar{\mathbf{X}}_n]|\F_{n-1}]=0=\Gamma_{N\theta}^{t_1 t_2}.$$

\textit{Computation of $\Gamma_{QQ}:=a.s.-\lim_n\E[\Delta\mathbf{M}_{Q,n}(\Delta\mathbf{M}_{Q,n})^\top|\F_{n-1}]$}.
It is immediate to see that for any $t_1\neq t_2$
$$\E[\Delta Q_{t_1,n}\Delta Q_{t_2,n}|\F_{n-1}]=-\mu_{n-1}(t_1)\mu_{n-1}(t_2)\stackrel{a.s.}{\longrightarrow}\Gamma_{QQ}^{t_1 t_2}:=-\mu(t_1)\mu(t_2),$$
while for $t_1=t_2=t$ we have
$$\E[\Delta Q_{t,n}^2|\F_{n-1}]=\mu_{n-1}(t)(1-\mu_{n-1}(t))\stackrel{a.s.}{\longrightarrow}\Gamma_{QQ}^{t t}:=\mu(t)(1-\mu(t)).$$

\textit{Remaining terms in $\Gamma$}.
Finally, we have that for any $t_1\neq t_2$,
$$\E[\Delta\mathbf{M}_{Z(t_1),n}\Delta M_{Q_{t_2},n}|\F_{n-1}]=
\E[\E[\Delta\mathbf{M}_{Z(t_1),n}|\F_{n-1},T_n]\Delta M_{Q_{t_2},n}|\F_{n-1}]=0=\Gamma_{ZQ}^{t_1 t_2},$$
$$\E[\Delta\mathbf{M}_{N_{t_1},n}\Delta M_{Q_{t_2},n}|\F_{n-1}]=
\E[\E[\Delta\mathbf{M}_{N_{t_1},n}|\F_{n-1},T_n]\Delta M_{Q_{t_2},n}|\F_{n-1}]=0=\Gamma_{NQ}^{t_1 t_2},$$
$$\E[\Delta\mathbf{M}_{\theta_{t_1},n}\Delta M_{Q_{t_2},n}|\F_{n-1}]=
\E[\E[\Delta\mathbf{M}_{\theta_{t_1},n}|\F_{n-1},T_n]\Delta M_{Q_{t_2},n}|\F_{n-1}]=0=\Gamma_{\theta Q}^{t_1 t_2}.$$

Since the assumptions are all satisfied, we can apply the CLT of the SA to the dynamics~\eqref{eq:dynamics_proofs_6_strata},
so obtaining a Gaussian asymptotic distribution for the process $\{\mathbf{W}_n;n\geq1\}$, with asymptotic variance
$$\Sigma\ :=\
\int_0^{\infty}e^{u(\frac{\mathbf{I}}{2}-\mathcal{D}f_{W}(\mathbf{W}))}
\Gamma e^{u(\frac{\mathbf{I}}{2}-\mathcal{D}f_{W}(\mathbf{W}))^\top}du.$$
This concludes the proof.
\endproof

\proof[Proof of Theorem~\ref{thm:second_order_result_adjusted}]
The structure of this proof is analogous to the proof of Theorem~\ref{thm:second_order_result_strata}.
In particular, we initially need to express in the SA form~\eqref{SAP} the joint dynamics of the following processes:
\begin{itemize}
\item[(1)] the urn proportion in correspondence of all the covariate profiles,
$$\mathbf{Z}_n:=(\mathbf{Z}_n(t),t\in\tau)^\top;$$
\item[(2)] the proportion of subjects assigned to the treatments in the study, $\tilde{\mathbf{N}}_{n}:=\mathbf{N}_{n}/n$;
\item[(3)] the adaptive estimators of features of interest related with the family of response distributions conditioned on
the covariates, $\hat{\mathbf{\beta}}_{n}:=(\hat{\mathbf{\beta}}^j_{n},j\in\{1,..,d\})^\top$.
\end{itemize}
Then, the CLT follows by applying the standard theory of the SA to the joint dynamics.

Using analogous arguments to the proof of Theorem~\ref{thm:second_order_result_strata},
we can obtain from~\eqref{eq:urn_dynamics} and~\eqref{SAP_theta_adjusted} the following joint dynamics:
\begin{equation}\label{eq:dynamics_proofs_4_adjusted}
\left\{\begin{aligned}
&\Delta\mathbf{Z}_n(t)\ =\ -r_n(\mathbf{Z}_{n-1}(t)-D_n(t)\mathbf{X}_n(t)),\\
&\Delta\tilde{\mathbf{N}}_{n}\ =\
-r_n\left(\tilde{\mathbf{N}}_{n-1}-\bar{\mathbf{X}}_n\right)+r_n\mathbf{R}_{N_t,n},\\
&\Delta\hat{\mathbf{\beta}}^j_{n}\ =\ -r_n\frac{\bar{X}_n^j}{\tilde{N}^j_{n-1}}
(f_{j}(\hat{\mathbf{\beta}}^j_{n-1})-\Delta \mathbf{M}_{j,n})+r_n\mathbf{R}_{\beta^j,n},
\end{aligned}\right.\end{equation}
where $\mathbf{R}_{N_t,n},\mathbf{R}_{\beta^j,n}\in{\mathcal F}_{n}$ are
suitable random variables that converges to zero a.s. and
$(\E[\|\mathbf{R}_{N_t,n}\|^2]+\E[\|\mathbf{R}_{\beta^j,n}\|^2])=o(n)$.
Now, in order to express the dynamics in~\eqref{eq:dynamics_proofs_4_adjusted} in the SA form~\eqref{SAP},
we need to rewrite it as follows:
\begin{equation}\label{eq:dynamics_proofs_5_adjusted}
\left\{\begin{aligned}
&\Delta\mathbf{Z}_n(t)\ =\ -r_nf_{Z,t}(\mathbf{Z}_{n-1}(t))\ +\ r_n\Delta \mathbf{M}_{Z(t),n},\\
&\Delta\tilde{\mathbf{N}}_{n}\
-r_nf_{N}(\mathbf{Z}_{n-1},\tilde{\mathbf{N}}_{n-1},\hat{\beta}_{n-1})\ +\
r_n(\Delta \mathbf{M}_{N,n}+\mathbf{R}_{N,n}),\\
&\Delta\hat{\mathbf{\beta}}^j_{n}\ =\ -r_n
f_{\beta^j}(\mathbf{Z}_{n-1}(t),\tilde{\mathbf{N}}_{n-1},\hat{\beta}^j_{n-1})
+r_n(\Delta \mathbf{M}_{\beta^j,n}+\mathbf{R}_{\beta^j,n}),
\end{aligned}\right.\end{equation}
where
\[\begin{aligned}
&f_{Z,t}(\mathbf{Z}_{n-1}(t)):=(I-H(t))\mathbf{Z}_{n-1}(t)+\mathbf{v}(t)(\mathbf{1}^\top\mathbf{Z}_{n-1}(t)-1),\\
&f_{N}(\mathbf{Z}_{n-1},\tilde{\mathbf{N}}_{n-1},\hat{\beta}_{n-1}):=
    \left(\tilde{\mathbf{N}}_{n-1}-\sum_{s=1}^K\mu_{n-1}(s)\mathbf{Z}_{n-1}(s)\right),\\
&f_{\beta^j}(\mathbf{Z}_{n-1},\tilde{\mathbf{N}}_{n-1},\hat{\beta}^j_{n-1}):=
\frac{\sum_{s=1}^K\mu_{n-1}(s)\mathbf{Z}_{n-1}(s)}{\tilde{N}^j_{n-1}}f_{j}(\hat{\mathbf{\beta}}^j_{n-1}),
\end{aligned}\]
and
\[\begin{aligned}
\Delta \mathbf{M}_{Z(t),n}&&:=&(D_n\mathbf{X}_n-H\mathbf{Z}_{n-1})(t),\\
\Delta \mathbf{M}_{N,n}\ &&:=&\ (\bar{\mathbf{X}}_n-\sum_{s=1}^K\mu_{n-1}(s)\mathbf{Z}_{n-1}(s)),\\
\Delta \mathbf{M}_{\beta^j,n}\ &&:=&\ (\bar{\mathbf{X}}_n-\sum_{s=1}^K\mu_{n-1}(s)\mathbf{Z}_{n-1}(s))
    \frac{f_{j}(\hat{\mathbf{\beta}}^j_{n-1})}{\tilde{N}^j_{n-1}} +\ \frac{\bar{X}_n^j}{\tilde{N}^j_{n-1}}\Delta \mathbf{M}_{j,n},
\end{aligned}\]
are martingale increments since $\E[D_n(t)|T_n,\mathbf{\bar{X}}_n]=H(t)$,
$\E[\bar{\mathbf{X}}_n|\mathcal{F}_{n-1}]=\mathbf{Z}_{n-1}(T_n)$,
\break
$\E[\Delta \mathbf{M}_{j,n}|\mathcal{F}_{n-1},T_n,\bar{\mathbf{X}}_n]=0$,
$\E[\ind_{\{T_n=t\}}|\mathcal{F}_{n-1}]=\mu_{n-1}(t)$.

Let us now introduce the joint processes $\{\mathbf{W}_n,n\geq1\}$ defined as
$\mathbf{W}_n:=(\mathbf{Z}_n,\tilde{\mathbf{N}}_{n},\hat{\mathbf{\beta}})^\top$,
and note that its dynamics can be expressed in the SA form~\eqref{SAP} as follows:
\begin{equation}\label{eq:dynamics_proofs_6_adjusted}
\Delta\mathbf{W}_n\ =\ -r_nf_{W}(\mathbf{W}_{n-1})\ +\ r_n(\Delta\mathbf{M}_{W,n}+\mathbf{R}_{W,n}),
\end{equation}
where
\begin{itemize}
\item[(i)] $f_{W}:=(f_{Z},f_{N},f_{\beta})^\top$, where $f_{Z}:=(f_{Z,t},t\in\tau)^\top$ and
$f_{\beta}:=(f_{\beta^j},j\in\{1,..,d\})^\top$;
\item[(ii)] $\Delta\mathbf{M}_{W,n}:=(\Delta\mathbf{M}_{Z,n},\Delta\mathbf{M}_{N,n},\Delta\mathbf{M}_{\beta,n})^\top$,
where $\Delta\mathbf{M}_{Z,n}:=(\Delta\mathbf{M}_{Z(t),n},t\in\tau)^\top$ and
$\Delta\mathbf{M}_{\beta,n}:=(\Delta\mathbf{M}_{\beta^j,n}j\in\{1,..,d\})^\top$;
\item[(iii)] $\mathbf{R}_{W,n}:=(\mathbf{0},\mathbf{R}_{N,n},\mathbf{R}_{\beta,n})^\top$, where
$\mathbf{R}_{\beta,n}:=(\mathbf{R}_{\beta^j,n},j\in\{1,..,d\})^\top$.
\end{itemize}
Since $\mathbf{R}_{W,n}\stackrel{a.s.}{\longrightarrow} \mathbf{0}$ and, using~\eqref{ass:martingale_theta_finite}
in~\ref{assum:6b}, $\sup_n\E[\|\Delta\mathbf{M}_{W,n}\|^2]<\infty$,
we have that the set $\mathcal{W}$ of the limiting values of $\mathbf{W}_n$ is a set stable by the flow of $\dot{\mathbf{W}}=-f_W(\mathbf{W})$.
Notice that by~\ref{assum:7b} the set $\{\mathbf{w}:f_W(\mathbf{w})=\mathbf{0}\}$ is composed only by the element
$\mathbf{W}:=(\mathbf{v},\mathbf{x}_0,\mathbf{\beta})^\top$, where
$\mathbf{v}:=(\mathbf{v}(t),t\in\tau)^\top$.
Moreover, by~\ref{assum:6b}
we have $\hat{\mathbf{\beta}}_n\stackrel{a.s.}{\longrightarrow}\mathbf{\beta}$,
and from Theorem~\ref{thm:first_order_result} we have $\mathbf{Z}_n\stackrel{a.s.}{\longrightarrow}\mathbf{v}$ and $\tilde{\mathbf{N}}_{n}\stackrel{a.s.}{\longrightarrow}
\mathbf{x}_0=\sum_{s=1}^Kf_{\mu,s}(\mathbf{x}_0,\mathbf{\beta})\mathbf{v}(s)=\sum_{s=1}^K\mu(s)\mathbf{v}(s)$,
which implies $\mathbf{W}_n\stackrel{a.s.}{\longrightarrow}\mathbf{W}$.

In order to show the existence of a stable attracting area which contains a neighborhood of $\mathbf{W}$, 
it is sufficient (see \cite[p. 1077]{FortPages}) to show that $\{\Re e(Sp(\mathcal{D}f_{W}(\mathbf{W})))>0\}$, where
\begin{equation}\label{eq:matrix_Df_proof_adjusted}
\mathcal{D}f_{W}(\mathbf{W})=
\begin{pmatrix}
\mathcal{D}_Zf_{Z}(\mathbf{W}) & 0 & 0 \\
\mathcal{D}_Zf_{N}(\mathbf{W}) & \mathcal{D}_Nf_{N}(\mathbf{W}) & \mathcal{D}_{\beta}f_{N}(\mathbf{W}) \\
0 & 0 & \mathcal{D}_{\beta}f_{\beta}(\mathbf{W})
\end{pmatrix},
\end{equation}
and
\begin{itemize}
\item[(i)] $\mathcal{D}_Zf_{Z}(\mathbf{W})$ is a block-diagonal matrix, whose $t^{th}$ block is\\
$[\mathcal{D}_Zf_{Z}(\mathbf{W})]^{tt}=(I-H(t)+\mathbf{v}(t)\mathbf{1}^\top)$;
\item[(ii)] $\mathcal{D}_{Z}f_{N}(\mathbf{W}):=-(\mu(1)I,..,\mu(K)I)$;
\item[(iii)] $\mathcal{D}_{N}f_{N}(\mathbf{W}):=I-\sum_{s=1}^K\mathbf{v}(s)\mathcal{D}_{N}f_{\mu,s}(\mathbf{W})^\top$;
\item[(iv)] $\mathcal{D}_{\beta}f_{N}(\mathbf{W}):=-\sum_{s=1}^K\mathbf{v}(s)\mathcal{D}_{\beta}f_{\mu,s}(\mathbf{W})^\top$;
\item[(v)] $\mathcal{D}_{\beta}f_{\beta}(\mathbf{W})$ is a block-diagonal matrix, whose $j^{th}$ block is\\
$[\mathcal{D}_{\beta}f_{\beta}(\mathbf{W})]^{jj}=\mathcal{D}f_{j}(\mathbf{\beta}^j)$.
\end{itemize}
Note from the structure of $\mathcal{D}f_{W}(\mathbf{W})$ in~\eqref{eq:matrix_Df_proof_adjusted} that
$\{\Re e(Sp(\mathcal{D}f_{W}(\mathbf{W})))>0\}$ follows by establishing that for any $t\in\tau$ and $j\in\{1,..,d\}$
$$\{\Re e(Sp(I-H(t)+\mathbf{v}(t)\mathbf{1}^\top))>0\},\ \mbox{ and }\ \{\Re e(Sp(\mathcal{D}f_{j}(\mathbf{\beta}^j)))>0\},$$
and $\{\Re e(Sp(\sum_{s=1}^K\mathbf{v}(s)\mathcal{D}_{N}f_{\mu,s}(\mathbf{W})^\top))<1\}.$
Analogous to the proof of Theorem~\ref{thm:second_order_result_strata},
the first two conditions follow from~\ref{assum:5} and~\ref{assum:6b}, respectively,
while the last condition follows from~\ref{assum:7b}.

We now show that the assumptions of the CLT for processes in the SA form are satisfied by
the dynamics in~\eqref{eq:dynamics_proofs_6_adjusted} of the joint process $\{\mathbf{W}_{n},n\geq1\}$.
First, note that using the above arguments we obtain $\{\Re e(Sp(\mathcal{D}f_{W}(\mathbf{W})))>1/2\}$.
Then, it is immediate to see that $\E[\|\mathbf{R}_{W,n}\|^2]=o(n)$ and
the first condition in~\eqref{HypDM} is satisfied using~\eqref{ass:martingale_theta_finite} in Assumption~\ref{assum:6b}.
Concerning the second condition in~\eqref{HypDM}, we need to show that there exists a deterministic symmetric positive
semidefinite matrix $\Gamma$ such that
$$\E[\Delta\mathbf{M}_{W,n}(\Delta\mathbf{M}_{W,n})^\top|\F_{n-1}]\stackrel{a.s.}{\longrightarrow}\Gamma=
\begin{pmatrix}
\Gamma_{ZZ} & \Gamma_{ZN} & \Gamma_{Z\beta}\\
\Gamma_{ZN}^\top & \Gamma_{NN} & \Gamma_{N\beta}\\
\Gamma_{Z\beta}^\top & \Gamma_{N\beta}^\top & \Gamma_{\beta \beta}
\end{pmatrix}.$$
As in the proof of Theorem~\ref{thm:second_order_result_strata}, note that since
$f_{j}(\hat{\mathbf{\beta}}^j_{n-1})\stackrel{a.s.}{\longrightarrow}0$,
this term does not contribute to $\Gamma$; hence in the following calculations
they will be omitted by $\Delta \mathbf{M}_{\beta^j,n}$.
We now proceed with the computation of the terms in $\Gamma$.
The calculations that follow by~\eqref{eq:convergence_X_g} are here omitted since
they are analogous in the proof of Theorem~\ref{thm:second_order_result_strata}.

\textit{Computation of $\Gamma_{ZZ}:=a.s.-\lim_n\E[\Delta\mathbf{M}_{Z,n}(\Delta\mathbf{M}_{Z,n})^\top|\F_{n-1}]$}.
For any $t_1,t_2\in\tau$
$$\Gamma_{ZZ}^{t_1t_2}:=\E[D(t_1)\mathbf{g}(t_1,T,\bar{\mathbf{X}})\mathbf{g}^\top(t_2,T,\bar{\mathbf{X}})D^\top(t_2)]-\mathbf{v}(t_1)\mathbf{v}^\top(t_2),$$
where $\mathbf{g}\in\mathcal{S}$ is a $d$-multivariate function defined in~\eqref{def:vector_g}.

\textit{Computation of $\Gamma_{NN}:=a.s.-\lim_n\E[\Delta\mathbf{M}_{N,n}(\Delta\mathbf{M}_{N,n})^\top|\F_{n-1}]$}.
Note that
\[\begin{aligned}
&\E[\Delta\mathbf{M}_{N,n}(\Delta\mathbf{M}_{N,n})^\top|\F_{n-1}]\\
&=diag\left(\sum_{s=1}^K\mu_{n-1}(s)\mathbf{Z}_{n-1}(s)\right)-\left(\sum_{s=1}^K\mu_{n-1}(s)\mathbf{Z}_{n-1}(s)\right)
\left(\sum_{s=1}^K\mu_{n-1}(s)\mathbf{Z}_{n-1}(s)\right)^\top\\
&\stackrel{a.s.}{\longrightarrow}\Gamma_{NN}:=diag\left(\sum_{s=1}^K\mu(s)\mathbf{v}(s)\right)-\left(\sum_{s=1}^K\mu(s)\mathbf{v}(s)\right)
\left(\sum_{s=1}^K\mu(s)\mathbf{v}(s)\right)^\top.
\end{aligned}\]

\textit{Computation of $\Gamma_{ZN}:=a.s.-\lim_n\E[\Delta\mathbf{M}_{Z,n}(\Delta\mathbf{M}_{N,n})^\top|\F_{n-1}]$}.
For any $t\in\tau$
\[\begin{aligned}
&\E[\Delta\mathbf{M}_{Z(t),n}(\Delta\mathbf{M}_{N,n})^\top|\F_{n-1}]
=\E[D_n(t)\mathbf{X}_n(t)(\Delta\mathbf{M}_{N,n})^\top|\F_{n-1}]\\
&=\sum_{s=1}^K\mu_{n-1}(s)
\left(\E[D_n(t)\mathbf{X}_n(t)\bar{\mathbf{X}}_n^\top|\F_{n-1},T_n=s]-H(t)\mathbf{Z}_{n-1}(t)\mathbf{Z}_{n-1}^\top(s)\right)\\
&\stackrel{a.s.}{\longrightarrow}\Gamma_{ZN}^{t}:=\sum_{s=1}^K\mu(s)
\left(H(t_1)G(t_1,s)diag(\mathbf{v}(s))-\mathbf{v}(t_1)\mathbf{v}^\top(s)\right),
\end{aligned}\]
where $G(t_1,s)$ is a matrix whose columns are $\{\mathbf{g}(t_1,s,\mathbf{e}_j);j\in\{1,..,d\}\}$.

\textit{Computation of $\Gamma_{\beta\beta}:=a.s.-\lim_n\E[\Delta\mathbf{M}_{\beta,n}(\Delta\mathbf{M}_{\beta,n})^\top|\F_{n-1}]$}.
Since for any $j_1\neq j_2$ we have $\E[\Delta\mathbf{M}_{\beta^{j_1},n}(\Delta\mathbf{M}_{\beta^{j_2},n})^\top|\F_{n-1}]=0$,
$\Gamma_{\beta\beta}$ is a block-diagonal matrix. In particular, for any $j\in\{1,..,d\}$ we have
\[\begin{aligned}
&\E[\Delta\mathbf{M}_{\beta^j,n}(\Delta\mathbf{M}_{\beta^j,n})^\top|\F_{n-1}]=\\
&(\tilde{N}^j_{n-1})^{-2}\left(\sum_{s=1}^K\mu_{i-1}(s)Z^j_{i-1}(s)\right)
\E[\Delta\mathbf{M}_{j,n}(\Delta\mathbf{M}_{j,n})^\top|\F_{n-1},\bar{X}_n^j=1]\\
&\stackrel{a.s.}{\longrightarrow}\Gamma_{\beta\beta}^{jj}:=
\left(\sum_{s=1}^K\mu(s)v^j(s)\right)^{-1}\E[\Delta\mathbf{M}_{j}(\Delta\mathbf{M}_{j})^\top|\bar{X}^j=1].
\end{aligned}\]

\textit{Computation of $\Gamma_{Z\beta}:=a.s.-\lim_n\E[\Delta\mathbf{M}_{Z,n}(\Delta\mathbf{M}_{\beta,n})^\top|\F_{n-1}]$}.
For any $t\in\tau$ and $j\in\{1,..,d\}$, we have that
\[\begin{aligned}
&\E[\Delta\mathbf{M}_{Z(t),n}(\Delta\mathbf{M}_{\beta^j,n})^\top|\F_{n-1}]
=\E[D_n(t)\mathbf{X}_n(t)(\Delta\mathbf{M}_{\beta^j,n})^\top|\F_{n-1}]\\
&=(\tilde{N}^j_{n-1})^{-1}(\sum_{s=1}^K\mu_{i-1}(s)Z^j_{i-1}(s))
\E[D_n(t)\mathbf{X}_n(t)(\Delta\mathbf{M}_{j,n})^\top|\F_{n-1},\bar{X}_n^j=1]\\
&\stackrel{a.s.}{\longrightarrow}\Gamma_{Z\beta}^{tj}:=\E[D(t)\mathbf{g}(t,T,\mathbf{e}_j)\Delta\mathbf{M}_{j}^\top|\bar{X}^j=1].
\end{aligned}\]

\textit{Computation of $\Gamma_{N\beta}:=a.s.-\lim_n\E[\Delta\mathbf{M}_{N,n}(\Delta\mathbf{M}_{\beta,n})^\top|\F_{n-1}]$}.
It can immediately be seen that, for any $j\in\{1,..,d\}$,
$$\E[\Delta\mathbf{M}_{N,n}(\Delta\mathbf{M}_{\beta^j,n})^\top|\F_{n-1}]=
\E[\Delta\mathbf{M}_{N,n}\E[\Delta\mathbf{M}_{\beta^j,n}|\F_{n-1},T_n,\bar{\mathbf{X}}_n]|\F_{n-1}]=0=\Gamma_{N\beta}^{j}.$$

Since the assumptions are all satisfied, we can apply the CLT of the SA to the dynamics~\eqref{eq:dynamics_proofs_6_adjusted},
so obtaining a Gaussian asymptotic distribution for the process $\{\mathbf{W}_n;n\geq1\}$, with asymptotic variance
$$\Sigma\ :=\
\int_0^{\infty}e^{u(\frac{\mathbf{I}}{2}-\mathcal{D}f_{W}(\mathbf{W}))}
\Gamma e^{u(\frac{\mathbf{I}}{2}-\mathcal{D}f_{W}(\mathbf{W}))^\top}du.$$
This concludes the proof.
\endproof


\end{document}